\documentclass{article}
\usepackage{array,booktabs}
\usepackage{bigstrut}
\makeatletter
\newlength\mylena
\newlength\mylenb
\newcommand\mystrut[1][2]{%
    \setlength\mylena{#1\ht\@arstrutbox}%
    \setlength\mylenb{#1\dp\@arstrutbox}%
    \rule{0pt}{4mm}}
\makeatother

\usepackage{fullpage}
\newcounter{thm}

\newtheorem{definition}[thm]{Definition}

\usepackage{times}

\newcommand{\Diadot}{\raisebox{0.1em}{\rotatebox[origin=c]{45}{$\textstyle\boxdot$}}}

\def \bseq {\mathop{\mathrm{bseq}}}

\newcommand{\etal}{et al.}
\usepackage{mine}
\usepackage{yhmath}
\usepackage{timestamp}
\usepackage{oz}
\usepackage{zed-csp}
\usepackage{rcp} 
\usepackage{graphicx}
\usepackage{color}
\usepackage{stmaryrd}
\usepackage{footnote}

\usepackage{amsmath}
\usepackage{mathrsfs}
\usepackage{rcp}
\usepackage{mine}
\usepackage{color}		
\usepackage{epsfig}
\usepackage{oz}
\usepackage{zed-csp}
\usepackage{timestamp}

\let \refeqn \refeq

\renewcommand{\qed}{\ensuremath{{}_\Box}}

\newcommand{\st}{~{\scriptscriptstyle ^\bullet}~}

\def \rely {\mathop{\textsc{Rely}}}

\newcommand{\NoteEnv}[3]{\newenvironment{#1}{\par\color{#3}#2: }{}}
\definecolor{brijeshcolor}{rgb}{0,0,1}
\definecolor{johncolor}{cmyk}{1,0.3,0.4,0.3}

\NoteEnv{brijesh}{Brijesh}{brijeshcolor}
\NoteEnv{john}{John}{johncolor}

\def\abssynt{\mathop{:\joinrel:\joinrel=}}

\newcommand{\Par}{\textstyle\mathop{\|}}

\def\ch{\mathbin{;}}

\def \bs {\backslash}

\def \dom {\mathrm{dom}}
\def \ran {\mathrm{ran}}

\def \seq {\mathrm{seq}}

\newcommand{\tott}{\fontfamily{txtt}\selectfont}

\newcommand{\Add}{{\tt add} }
\newcommand{\Rem}{{\tt remove} }
\newcommand{\Cont}{{\tt contains} }

\date{}

\begin{document}

\title{Verifying linearizability: A comparative survey}
\author{Brijesh Dongol,  
  Brunel University UK \\[2mm]
  John Derrick,
  University of Sheffield, UK
}






\maketitle

\begin{abstract}
  Linearizability has become the key correctness criterion for
  concurrent data structures, ensuring that histories of the
  concurrent object under consideration are consistent, where
  consistency is judged with respect to a sequential history of a
  corresponding abstract data structure. Linearizability allows any
  order of concurrent (i.e., overlapping) calls to operations to be
  picked, but requires the real-time order of non-overlapping to be
  preserved. Over the years numerous techniques for verifying
  linearizability have been developed, using a variety of formal
  foundations such as refinement, shape analysis, reduction,
  etc. However, as the underlying framework, nomenclature and
  terminology for each method differs, it has become difficult for
  practitioners to judge the differences between each approach, and
  hence, judge the methodology most appropriate for the data structure
  at hand. We compare the major of methods used to verify
  linearizability, describe the main contribution of each method, and
  compare their advantages and limitations.
\end{abstract}

\section{Introduction}
\label{sec:introduction}

Highly optimised fine-grained concurrent algorithms are increasingly
being used to implement concurrent objects for modern multi/many-core
applications due to the performance advantages they provide over their
coarse-grained counterparts. Due to their complexity, correctness of
such algorithms is notoriously difficult to judge and formal
verification has uncovered subtle bugs in published algorithms that
were previously thought correct \cite{Doherty03,CG05}. The main
correctness criterion for concurrent algorithms is
\emph{linearizability}, which defines consistency conditions on the
history of invocation and response events generated by an execution of
the algorithm at hand \cite{Herlihy90}. Linearizability requires every
operation call to take effect at some point between its invocation and
response events. Concurrent calls may take effect in any order, but
the real-time order of sequential operation calls must be
maintained. A (concurrent) history is linearizable iff there is some
order for the effects of the operation calls that corresponds to a
valid sequential history, and a concurrent object is linearizable iff
each of its histories is linearizable.

With the increasing demand for high performance in modern computing
systems, there has been an increase in the sophistication in the
algorithms implementing linearizable concurrent objects. The subtle
nature of their behaviours has meant that such algorithms must be
proved correct. However, scalability of the proof methods remains an
open problem, and hence, an immense amount of research effort has been
devoted to linearizability verification. Unfortunately, each new
method uses specialised formal frameworks (with their own
specification languages with associated semantics), making it
difficult to judge the merits of each approach. In this paper, we
present a comparative survey of the major techniques that have been
developed to examine the advantages and downfalls of each. We aim to
make our comparison comprehensive, but with the scale of development
in this area, it is inevitable that some published methods for
linearizability verification will be left out. The survey does not aim
to be comprehensive about fine-grained algorithms, nor about the sorts
of properties that these algorithms possess; for this,
\cite{HS08,MS04} are already excellent resources. Instead, this survey
is aimed at improving ones understanding of the fundamental challenges
of linearizability verification and identifying avenues of future
work. Some questions to be asked about the different methods
are:\vspace{-1.5mm}
\begin{itemize}
\item \emph{Locality of the proof method.} How is a proof of
  linearizability (a global property) decomposed so that proofs are
  performed in a process-local manner?
\item \emph{Compositionality of the proof method.} Does the method
  support compositional proofs, where interference is captured
  abstractly?
\item \emph{Contribution of the framework.} Does the underlying
  framework contribute to simpler proofs? If so, how?
\item \emph{Algorithms verified.} Which algorithms have been verified
  and how complex are these algorithms?
\item \emph{Mechanisation.} Has the method been mechanised? If so,
  what is the level of automation?
\item \emph{Completeness.} We say a method is \emph{complete} if
  whenever an implementation is linearizable with respect to an
  abstract specification, it can be proved linearizable using the
  method. Has completeness of the proof methods at hand been shown? If
  not, what is the verification power of each method?
\end{itemize}\vspace{-1.5mm}

Linearizability is a condition on concurrent objects, where
consistency of an object in question judged with respect to a
sequential counterpart. Most techniques for verifying linearizability
involve identification of a \emph{linearization point} of each
operation, which is an atomic statement whose execution causes the
\emph{effect} of the operation to take place, i.e., executing a
linearization point has the same effect as executing the corresponding
abstract operation. The ordering of linearization points in a
concurrent object's trace defines a linearization of the history
corresponding to the trace.

It turns out that identification of linearization points is a
non-trivial task. Some algorithms have simple fixed linearization
points, while others have external linearization points that are
determined by the execution of other operations. For yet more complex
algorithms, each concrete state corresponds to several possible
abstract queues, and hence, the linearization points are dependent on
the possible future behaviours of the
program. 
We therefore consider three case studies for comparison with
increasing levels of difficulty --- (1) an \emph{optimistic set} with
operations \Add and {\tt remove}, both of which have fixed linearization
points (2) a \emph{lazy set} \cite{HHLMSS07}, which is the optimistic
set together with a wait-free \Cont operation that may be linearized
externally; and (3) Herlihy/Wing's array-based queue \cite{Herlihy90},
with future-dependant linearization points.  

This paper is structured as follows. In
\refsec{sec:linearizability-1}, we describe linearizability, present
its original definition using the nomenclature of Herlihy/Wing. In
\refsec{sec:verify-line}, we present an overview of the different
methods that have been developed for verifying linearizability, which
includes simulation, data refinement, auxiliary variables, shape
analysis, etc. Sections \ref{sec:case-study:-fine},
\ref{sec:case-study-2} and \ref{sec:case-study-3} present our case
studies, and \refsec{sec:conclusions} concludes the paper.

\section{Linearizability}
\label{sec:linearizability-1}

Concurrent objects allow different processes to concurrently execute
its operations by interleaving their atomic statements so that the
intervals of execution for different operation calls overlap. Such
objects are generally more efficient than coarse-grained counterparts,
however, what we mean by correctness is open to interpretation. The
most widely accepted correctness condition is \emph{linearizability}
\cite{Herlihy90}, which defines the meaning of a concurrent object by
mapping its (concurrent) histories to those of a (sequential)
specification. We refer to the implementation as the \emph{concrete
  object} and the specification as the \emph{abstract object}.

We motivate linearizability using a non-blocking stack
algorithm (\refsec{sec:exampl-treib-stack}) before presenting the
formal definition (\refsec{sec:linearizability}). In
\refsec{sec:line-observ-refin}, we discuss the correspondence between
linearizability and observational refinement. 

\subsection{Example: The Treiber stack}
\label{sec:exampl-treib-stack}
\reffig{fig:TS} presents a simple non-blocking stack example due to
Treiber \cite{Tre86}, which has become a standard case study from the
literature. The version we use assumes garbage collection to avoid the
so-called ABA problem \cite{Doherty03}, where changes to shared
pointers to undetected due to the value changing from $A$ to $B$, but
then back to $A$. Without garbage collection, solving this requires
additional complexities such as version numbers for pointers to be
introduced; such details are elided in this paper.
\begin{figure}[h]
\rule{\textwidth}{1pt}

  \centering 
  \begin{minipage}[t]{0.9\columnwidth}
    {\tt \footnotesize Init: Head = null}\medskip

    \begin{sidebyside}
      \tt \footnotesize push(v)

      \ H1: n := \textbf{new}(Node);

      \ H2: n.val := v;

      \ H3: \textbf{repeat}

      \ H4: \ \ ss := Head;
    
      \ H5: \ \ n.next := ss;

      \ H6: \ \ \textbf{until} CAS(Head,ss,n)

      \ H7: \textbf{return} \nextside 

\tt \footnotesize pop: lv

      \ P1: \textbf{repeat }

      \ P2: \ \ ss := Head;

      \ P3: \ \ \textbf{if} ss = null \textbf{then}

      \ P4: \ \ \ \ \textbf{return} empty


      \ P5: \ \ ssn := ss.next;

      \ P6: \ \ lv := ss.val

      \ P7:\ \textbf{until} CAS(Head,ss,ssn);

      \ P8: \textbf{return} lv
    \end{sidebyside}
  \end{minipage}
\caption{The Treiber stack}
\label{fig:TS}
\rule{\textwidth}{1pt}
\end{figure}

The algorithm implements the abstract stack in
\reffig{fig:Abstract-TS}, where brackets `$\langle$' and `$\rangle$'
are used to explicitly define sequences, `$\cat$' denotes sequence
concatenation and `$\emptyseq$' denotes the empty sequence. The
abstract algorithm consists of a sequence of elements together with
two operations $push$ (that pushes its input $v \neq empty$ onto the
top of the stack) and $pop$ (that returns $empty$ and leaves the stack
unchanged if the stack is empty, and removes one element from the top
and returns the top element otherwise).

Such data structures (or more generally objects) are typically
realised as part of a system library, whose operations are thought of
as being invoked by \emph{client} processes. For reasoning purposes,
one typically thinks of an object as being executed by a \emph{most
  general client}, which formalises Herlihy and Wing's
\cite{Herlihy90} intuition that a process calls at most one operation
of the object at a time. For example, a most general client process of
a stack \cite{ARRSY07} is given in \reffig{fig:MGC}, where the {\tt ?}
test in the {\tt \textbf{if}}\, denotes non-deterministic
choice. Usage of a most general client for verification was, however,
proposed in much earlier work \cite{Doherty03}.

\begin{figure}[h]
  \rule{\textwidth}{1pt}

  \centering 
  \begin{minipage}[b]{0.48\columnwidth}
    {\tt \footnotesize Init: S = $\emptyseq$}\medskip

    \begin{minipage}[t]{0.48\columnwidth}
      \tt \footnotesize push(v)

      \ \textbf{atomic} \{
      
      \ \ \ \ S := $\langle$v$\rangle \cat$S
      
      \ \}
    \end{minipage}
    \begin{minipage}[t]{0.48\columnwidth}
      \tt \footnotesize pop: lv
      
      \ \textbf{atomic} \{
      
      \ \ \ \textbf{if} S = $\emptyseq$ \textbf{then}
      
        \ \ \ \ \ \textbf{return} empty

        \ \ \ \textbf{else}

        \ \ \ \ \ lv := head(S) ;
      
        \ \ \ \ \ S := tail(S) ;

        \ \ \ \ \ \textbf{return} lv \}
      \end{minipage}
    \caption{An abstract stack specification}
    \label{fig:Abstract-TS}
  \end{minipage}
  \hfill 
  \begin{minipage}[b]{0.45\columnwidth}
    \centering
    \begin{minipage}[b]{0.6\linewidth}
      {\tt \footnotesize client(Stack *st) \{

        \ \textbf{do} \{

        \ \ \ \textbf{if} (?)

        \ \ \ \ \ push(st, rand());

        \ \ \ \textbf{else}

        \ \ \ \ \ pop(st);

        \ \ \ \} \textbf{while} true;

        \}}
    \end{minipage}
  \caption{Most general client process for a Stack}
    \label{fig:MGC}
  \end{minipage}

\rule{\textwidth}{1pt}

\end{figure}

The implementation (\reffig{fig:TS}) has fine-grained
atomicity. Synchronisation is achieved using an atomic
compare-and-swap ($CAS$) operation:

\begin{minipage}[t]{\columnwidth}
  \tt CAS(gv, lv, nv) $\sdef$ atomic \{
  \begin{tabular}[t]{@{}l@{}}
    \tt if (gv = lv) then gv := nv ; return true \\
    \tt else return
    false \}
  \end{tabular}
\end{minipage}
which takes as input a (shared) variable {\tt gv}, an expected value
{\tt lv} and a new value {\tt nv}. In a single atomic step, the $CAS$
operation compares {\tt gv} to {\tt lv}, potentially updates {\tt gv}
to {\tt nv} and returns a boolean. In particular, if {\tt gv = lv}, it
updates {\tt gv} to {\tt nv} then returns true (to indicate that the
update was successful), otherwise it leaves everything unchanged then
returns false. The $CAS$ instruction is natively supported by most
mainstream hardware architectures. Operations that use $CAS$ typically
have a try-retry structure with a loop that stores (shared variable)
{\tt gv} locally in {\tt lv}, performes some calculations on {\tt lv}
to obtain {\tt nv} (a new value for {\tt gv}), then uses a $CAS$ to
attempt an update to {\tt gv}. If the $CAS$ fails, there must have
been some interference on {\tt gv} since it was stored locally at the
start of the loop, and in this case the operation retries by
re-reading {\tt gv}.

We now explain the (concrete) program in \reffig{fig:TS}, whose
operations both have the try-retry structure explained above. The
concrete $push$ operation first creates a new node with the value to
be pushed onto the stack ({\tt H1} and {\tt H2}). It then repeatedly
sets a local variable $ss$ to $Head$ ({\tt H4}) and the pointer of the
new node to $ss$ ({\tt H5}) until the $CAS$ succeeds ({\tt H6}), which
means $Head$ (still) equalled $ss$ and has atomically been set to the
new node $n$ ({\tt H6}). Note that the $CAS$ in $push$ does not
necessarily succeed: in case of a concurrent $push$ or $pop$
operation, $Head$ might have been changed between taking the snapshot
of $Head$ at {\tt H4} and execution of the $CAS$ at {\tt H6}. The
concrete $pop$ operation has a similar structure: it records the value
of $Head$ in $ss$ ({\tt P2}), and returns empty if $ss = null$ ({\tt
  P4}). Otherwise, the next node is stored in $ssn$ ({\tt P5}), the
return value is stored in $lv$ ({\tt P6}), and a $CAS$ is executed to
attempt to update $Head$ ({\tt P7}). If this $CAS$ succeeds, the $pop$
takes effect and the output value $lv$ is returned ({\tt P8}),
otherwise, $pop$ repeats its steps loading a new value of $Head$.



Linearizability ensures every (potentially concurrent) execution of
the implementation object (e.g., \reffig{fig:TS}) can be explained by
a sequential execution of a corresponding abstract object (e.g.,
\reffig{fig:Abstract-TS}). The sequential ordering is determined by
the order of \emph{linearization points} in the concurrent execution.
For the $push$ operation, a successful execution of the $CAS$ is a
linearization point as at this transition that adds element on to the
top of the stack. The $pop$ operation has two linearization points
depending on the value returned: if the stack is empty, the
linearization point is the statement labelled {\tt P2}, when $Head =
null$ is read, otherwise, the linearization point is a successful
execution of {\tt P7}. Note that {\tt P3} is not a linearization point
for an empty stack as the test only checks \emph{local} variable $ss$
--- the global variable $Head$ might be non-null again at this
point. Notice, also, that this example illustrates the fact different
statements may qualify as a linerization point depending on the actual
values returned by the operation. In the $pop$ operation, the location
of the linearization point depends on whether or not the stack is
empty.

\begin{figure}[h]
  \rule{\textwidth}{1pt} \smallskip

  \centering
  \scalebox{0.85}{\input{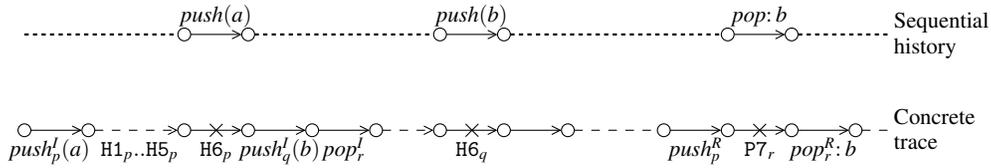}}
  \caption{Relating interleaved traces and linearizability}
  \label{fig:interleaved}
\rule{\textwidth}{1pt}
\end{figure}

A possible execution of the Treiber Stack (by a most general client)
is given in \reffig{fig:interleaved}, which depicts invocation (e.g.,
$push_p^I(b)$), response (e.g., $push_p^R$), and internal transitions
of operations $push_p(a)$, $push_q(b)$ and $pop_r:b$, where $p, q, r$
are processes. A cross on a transition arrow is used to denote the
linearization points. Although the three operations execute
concurrently by interleaving their statements, the order or
linearization points allows one to determine a sequential order for
the operations. Importantly, this order conforms to a valid execution
of the stack from \reffig{fig:Abstract-TS}.

\subsection{Formalising linearizability}
\label{sec:linearizability}

Although we have motivated our discussion of linearizability in terms
of the order of linearization points, and these being consistent with
an abstract counterpart, we have to relate this view to what is
observable in a program. In particular, what is taken to be observable
are the histories, which are sequences of invocation and response
events of operation calls on an object. This represents the
interaction between an object and its client via the object's external
interface. Thus, in \reffig{fig:interleaved}, the internal transitions
(including linearization points) are not observable.

Each observable event records the calling process (of type $P$), the
operation that is executed (of type $O$), and any input/output
parameters of the event (of type $V$).  Thus, we define
\cite{DSW11TOPLAS}:\vspace{1mm}

\hfill $Event  :\!:\!=  inv\lang P \times O \times V\rang \mid ret\lang P \times
O \times V\rang$\hfill{}\vspace{1mm}

\noindent For brevity, we use notation $op_p^I(x)$ and $op_p^R : r$
for events $inv(p, op, x)$ and $ret(p, op, r)$, respectively, and use
$op_p^I$ and $op_p^R$ to respectively denote invoke and return events
with no inputs or outputs. For an event $e = (p, op, x)$, we assume
the existence of projection functions $proc(e) = p$, $oper(e) = op$
and $par(e) = x$ that return the process, operation and input/output
parameter of event $e$, respectively. The definition of
linearizability is formalised in terms of the history of events, which
is represented formally by a sequence.  Let $\seq(X)$ denote sequences
of type $X$, which we assume are indexed from $0$ onward. A
\emph{history} is an element of $History \sdef \seq(Event)$, i.e., is
a sequence of events.

To motivate linearizability, consider the following history of a
concurrent stack, where execution starts with an empty
stack.\vspace{1mm}

\hfill $h_1 \sdef \aang{push_p^I(a), push_q^I(b), push_p^R, push_q^R}$ \hfill{}

\vspace{1mm}
\noindent Processes $1$ and $2$ are concurrent, and hence, the
operation calls may be linearized in either order, i.e., both
linearizations below are valid.  \vspace{1mm}

\hfill$hs_1  \sdef  \aang{push_p^I(a), push_p^R, push_q^I(b), push_q^R}
\quad\   
hs_2  \sdef  \aang{push_q^I(b), push_q^R, push_p^I(a), push_p^R}$\hfill{}

\vspace{1mm}
\noindent Assuming execution starts with an empty stack, the abstract
stack is $\aang{b, a}$ (with $b$ at the top) at the end of $hs_1$ and
$\aang{a, b}$ at the end of $hs_2$. Now suppose, history $h_1$ is
extended with a sequential $pop$ operation: \vspace{1mm}

\hfill
$h_2 \sdef h_1 \cat \aang{pop_r^I, pop_r^R:b}$ \hfill{}\vspace{1mm}

No linearization of $h_2$ may swap the order of the $pop$ with either
of the $push$ operations in $h_1$ because $pop^I_r$ occurs after the
return of both $push$ operation calls, i.e., their executions are not
concurrent. Furthermore, because elements must be inserted and removed
from a stack in a last-in-first-out order, adding the $pop$ that
returns $b$ restricts the valid linearizations of $h_2$. In
particular, the only sensible choice is one in which $push(b)$ occurs
after $push(a)$, i.e., \vspace{1mm}

\hfill$hs_3 \sdef hs_2 \cat \aang{pop_r^I,
  pop_r^R:b}$\hfill{}\vspace{1mm}
 
\noindent 
which results in an abstract stack $\aang{a}$ at the end of
execution. Sequential history $hs_1 \cat \aang{pop_r^I, pop_r^R:b}$ is
an invalid linearization of $h_2$. Now suppose $h_2$ is appended with
two more pop operations as follows: \vspace{1mm}

\hfill $h_3 \sdef h_2 \cat \aang{pop_s^I, pop_t^I, pop_s^R: a, pop_t^R:a}$ \hfill{}\vspace{1mm}

\noindent History $h_3$ cannot be linearized by any sequential stack
history --- the only possible stack at the end of $h_2$ is $\aang{a}$,
yet the additional events in $h_3$ are for two pop operations that are
successfully able to remove $a$ from the stack. A concurrent stack
that generates $h_3$ would therefore be deemed incorrect. Proving
linearizability of the Treiber stack ensures that a history such as
$h_3$ is never generated by the algorithm.

Formally, for $h \in History$, let $h\!|\!p$ denote the subsequence of
$h$ consisting of all invocations and responses of process $p$.  Two
histories $h_1$, $h_2$ are \emph{equivalent} if for all processes $p$,
$h_1\!|\!  p = h_2\!|\!p$.  An invocation $opi^I_p(x)$ \emph{matches}
a response $opj^R_q:y$ iff $opi = opj$ and $p = q$.  An invocation is
\emph{pending} in a history $h$ iff there is no matching response to
the invocation in $h$. We say the invocation is \emph{completed} in
$h$ iff it is not pending in $h$. We let $complete(h)$ denote the
maximal subsequence of history $h$ consisting of all (completed)
invocations and their matching responses in $h$, i.e., the history
obtained by removing all pending invocations within $h$.  For a
history $h$, let $<_h$ be an irreflexive partial order on operations,
where $opi <_h opj$ iff the response event of $opi$ occurs before the
invocation event of $opj$ in $h$.  A history $h$ is \emph{sequential}
iff the first element of $h$ is an invocation and each invocation
(except possibly the last) is immediately followed by its matching
response.

By assuming histories are  assumptions ensure
that the histories of the concurrent objects that one considers are
\emph{well-formed} \cite{Herlihy90}, i.e., for any history $h$ and
process $p$, the subhistory $h\!|\!p$ is sequential. For the rest of
this paper, we assume the objects in question are executed by a most
general client, and hence, that the histories in question are
well-formed.
\begin{definition}[Linearizability \cite{Herlihy90}]
  \label{def:linearizability}
  A history $hc$ is \emph{linearizable with respect to} a sequential
  history $hs$ iff $hc$ can be extended to a history $hc'$ by adding
  zero or more matching responses to pending invocations such that
  $complete(hc')$ is equivalent to $hs$ and ${}<_{hc} {}\subseteq
  {}<_{hs}{}$.
\end{definition}
We simply say $hc$ is linearizable if there exists a history $hs$ such
that $hc$ is linearizable with respect to $hs$.


Note that the definition of linearizability allows histories to be
extended with matching responses to pending invocations. This is
necessary because some operations may have passed their linearization
point, but not yet responded. For example, consider the following
history, where the stack is initially empty.
\begin{eqnarray}
  \label{eq:87}
  \langle push_p^I(x), pop_q^I, pop_q^R(x) \rangle
\end{eqnarray}
The linearization point of $push_p^I(x)$ has clearly been executed in
\refeq{eq:87} because $pop_q$ returns $x$, but \refeq{eq:87} is
incomplete because the $push_p$ is still pending. To cope with such
scenarios, by the definition of linearizability, \refeq{eq:87} may be
extended with a matching response to $push_p^I(x)$, and the extended
history mapped to the following sequential history:
$\langle push_p^I(x), push_p^R, pop_q^I, pop_q^R(x) \rangle$.

We have defined linearizability for concurrent histories. The purpose
of linearizability, however, is to define correctness of concurrent
objects with respect to some abstract specification. Thus, the
definition is lifted to the level of objects as follows.
\begin{definition} 
  A concurrent object is \emph{linearizable} with respect to a
  sequential abstract specification iff for any legal history $hc$
  of the concurrent object, there exists a sequential history $hs$
  of the abstract specification such that $hc$ is linearizable with
  respect to $hs$. 
\end{definition}

\subsection{Linearizability and observational refinement}
\label{sec:line-observ-refin}

A missing link in linearizability theory is the connection between
behaviours of objects and clients executing
together. Concurrent objects are
designed to satisfy specific correctness criterion defined by
concurrency theorists (e.g., sequential consistency, linearizability),
whereas programmers aim to understand conditions that enable an object
to be replaced by another.
Thus from a programmers perspective, one may ask: {\em How are the
  behaviours of a client that uses a sequential object $SO$ related to
  those of a client that uses a concurrent object $CO$ instead
  provided some correctness condition has been established between
  $CO$ and $SO$?}  An answer to this question was given by
Filipovi\'{c} \etal \cite{FORY10} who consider \emph{concurrent object
  systems} (which are collections of concurrent objects) and establish
a link between linearizability and observational refinement. Their
result covers data independent clients, i.e., those that communicate
only via their object systems and states that a concurrent object
system $COS$ \emph{observationally refines} a sequential object system
$AOS$ iff every object in $COS$ is \emph{sequentially consistent} with
respect to its corresponding object in $AOS$, where:
\begin{itemize}
\item $COS$ \emph{observationally refines} $AOS$ iff for any client
  program $P$ parameterised by an object system, the observable
  states\footnote{In their setting, the observable states consist of
    the variables of the clients only, i.e., none of the variables of
    the object system are observable.} of $P(COS)$ is a subset of the
  observable states of $P(AOS)$, i.e., $P(AOS)$ does not generate any
  new observations in comparison to $P(COS)$, and

\item $COS$ is \emph{sequentially consistent} with respect to $AOS$
  iff for every history $h_C$ of $COS$, there exists a sequential
  history $h_A$ such that the order of operation calls by the same
  process in $h_C$ is preserved in $h_A$.
\end{itemize}
It is well known that linearizability implies sequential consistency,
and hence, if $COS$ is linearizable with respect to $AOS$, then $COS$
also observationally refines $AOS$. Filipovi\'{c} \etal \cite{FORY10}
show that if the clients are able to communicate using at least one
additional shared variable, then $COS$ observationally refines $AOS$
iff $COS$ is linearizable with respect to $AOS$, where the definition
of linearizability is suitably generalised to object systems.

A further deficiency in linearizability theory is that assumes data
independence between libraries and clients, and hence only admits
pass-by-value parameter passing mechanisms. Real-world systems
however, also allow data sharing between libraries and clients, e.g.,
via pass-by-reference mechanisms. Here, ownership transfer between
shared resources may occur. To this end, Gotsman and Yang
\cite{GY12,GY13} have extended linearizability theory to cope
with parameter sharing between concurrent objects and its clients.
Cerone \etal \cite{CGY14} have further extended these results and
defined \emph{parameterised linearizability} that allows linearizable
objects to be taken as parameters to form more complex linearizable
objects.

Some authors have have presented a constructive methods for developing
fine-grained objects, dispensing with linearizability as a proof
obligation \cite{TW11,LiangFF12}. Instead, they focus on maintenance
of the observable behaviour of the abstract object directly. A survey
of techniques for verifying observational refinement lies outside the
scope of this paper. We focus on linearizability alone, and to this
end, an overview of different construction-based approaches to proving
linearizability is given in \refsec{sec:line-constr}.

\section{Verifying linearizability}
\label{sec:verify-line}



This section discusses linearizability verification in general. An
outline of different methods for decomposing proofs is given in
\refsec{sec:meth-proof-decomp} and \refsec{sec:degrees-difficulty}
describes how linearizability verification can be characterised based
on the linearization points. We give an overview of different methods
for verifying linearizability in Sections
\ref{sec:simul-with-canon}-\ref{sec:line-constr}.

\subsection{Methods for proof decomposition}
\label{sec:meth-proof-decomp}
Capturing the correspondence between a concurrent implementation
object and its sequential specification lies at the heart of
linearizability. It comes as no surprise therefore that almost all
methods for verifying linearizability involves some notion of
\emph{refinement} \cite{deRoever98} to link concrete and abstract
behaviours. In this section, characterise different algorithms based
on their linearization points, then review the different methods for
proving linearizability.

Typically, the internal representation of data in a concrete object
and its abstract specification differ, e.g., the Treiber stack is a
linked list (\reffig{fig:TS}), whereas its abstract specification is a
sequence of values (\reffig{fig:Abstract-TS}). A formal link between
their observable behaviours is given by \emph{data refinement}
\cite{deRoever98}, which uses a \emph{representation relation} to
relate concrete and abstract state spaces. Data refinement is a
system-wide (i.e., global) property and a monolithic proof of data
refinement quickly becomes unmanageable. Therefore, several methods
for decomposing it have been developed. The proof methods for
verifying linearizability all use some combination of the methods
below.

\begin{figure}[h]
\rule{\textwidth}{1pt}\medskip

  \centering
  \scalebox{0.6}{\input{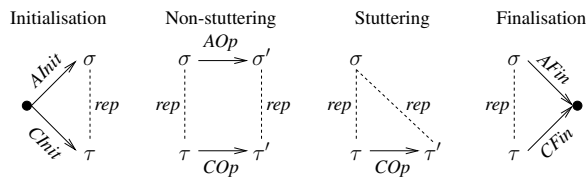}}
  \caption{Simulation diagrams}
  \label{fig:sim}
\rule{\textwidth}{1pt}
\end{figure}

\paragraph{Simulation} Decomposition of data refinement into
process-local proof obligations is achieved via \emph{simulation},
which allows one to reason about each transition of the concrete
object individually. \reffig{fig:sim} shows four typical simulation
rules where $AInit$, $AOp$ and $AFin$ are abstract initialisation,
operation and finalisation steps (and similarly $CInit$, $COp$ and
$CFin$), $\sigma$, $\sigma'$ are abstract states, $\tau$, $\tau'$ are
concrete states, and $rep$ is a representation relation between
abstract and concrete
states. 
Simulation proofs may be performed in a forwards or backwards manner
and although the set of diagrams for forwards and backward simulation
are the same, the order in which each diagram is traversed differs. It
turns out that neither forwards nor backwards simulation alone is
complete for verifying data refinement, but the combination of the two
forms a complete method \cite{deRoever98}.

\paragraph{Compositional frameworks} Compositional frameworks allow
one to reason about a concurrent program in a modular manner by
capturing the behaviour of the environment of a program abstractly as
a two-state relation \cite{deRoever01}.  For shared-variable
concurrency, a popular approach to compositionality is Jones'
rely-guarantee framework \cite{Jon83}, where a \emph{rely} condition,
states assumptions about a component's environment, and a
\emph{guarantee} condition describes the behaviour a component under
the assumption that the rely condition holds. A detailed survey of
different compositional verification techniques lies outside the scope
of this paper; we refer the interested reader to \cite{deRoever01}.

\paragraph{Reduction} Reduction enables one to ensure trace
equivalence of the fine-grained implementation and its coarse-grained
abstraction by verifying commutativity properties \cite{Lip75}. For
example, in a program $S1 ; S2$ if $S2$ performs purely local
modifications, $(S2_p \ch T_q) = (T_q ; S2_p)$ will hold for any
statement $T$ and processes $p$, $q$ such that $p \neq q$. Therefore,
$S1 ; S2$ in the program code may be treated as $atomic\{S1 ; S2\}$,
which in turn enables coarse-grained atomic blocks to be constructed
from finer-grained atomic statements in a manner that does not modify
the global behaviour of the algorithm. After a reduction-based
transformation, the remaining proof only needs to focus on verifying
linearizability of the coarse-grained abstraction
\cite{Gro08,Gro07,EQSST10}, which is simpler than verifying the
original program because fewer statements need to be considered.

\paragraph{Interval-based reasoning} Linearizability is a property
over the intervals in which operations execute, requiring a
linearization point to occur at some point between the operation's
invocation and response. Some methods use interval logics (for example
ITL \cite{Mos00,Mos97}) in order to exploit this. A program's
execution is treated as an interval predicate that defines the
evolution of the system over time, followed by a proof of abstraction
that projects concurrent behaviours to those of their sequential
counterparts.

\paragraph{Separation logic} Many linearizable objects are realised
using pointer-based structures such as linked lists. A well known
logic for reasoning about such implementations is separation logic
\cite{Rey02,ORY01}, which uses a so-called separating conjunction to
split the memory heap into disjoint portions, and reason about each of
these individually. Such techniques enable localised reasoning over
the part of the heap that is important for the assertions at hand. Of
course, linked lists are not the only application of separation logic;
for example, Gotsman and Yang \cite{GY13} use it to separate state
spaces of an object and its clients.

\medskip

The methods we discuss in this paper all use some combination of the
techniques above. Prior to exploring these methods in detail, we first
review the difficulties encountered when verifying linearizability.

\subsection{Difficulties in verifying linearizability}
\label{sec:degrees-difficulty}

Proving linearizability using linearization points is non-trivial, and
one may classify different types of algorithms based on their
linearization points (see \reftab{tab:degrees}). There are several
other algorithms that have not yet been formally verified correct, and
hence, the list of algorithms in \reftab{tab:degrees} is only
partial. The type of linearization point may be distinguished as being
\emph{fixed} (i.e,. the linearization point may be predetermined),
\emph{external} (i.e,. the execution of a different operation
potentially determines the linearization point) and
\emph{future-dependant} (i.e., the linearization point is determined
by considering future executions of the operation). Different
operations of the same object may have different types of linearization
points. In fact, even within an operation, there are different types
of linearization points depending on the value returned. For example,
the linearization points of the dequeue operation of the Michael/Scott
queue \cite{Michael96} has both future dependant (empty case) and
fixed (non-empty case) linearization points.

  \begin{table}[!h]
\rule{\textwidth}{1pt}
    \centering
      \begin{tabular}{>{\bfseries}p{40mm}| 
                         >{}p{20mm}
                         >{\arraybackslash}p{70mm}}
        Example algorithms & \textbf{Reference} & \textbf{Operations (linearization type)} \\
        \hline\mystrut
        Treiber stack & \cite{Tre86} &  Push (fixed), Pop (fixed) 
        \\\mystrut
        MS queue & \cite{Michael96} &    
        \begin{tabular}[t]{@{}l@{}}
          Enqueue (fixed), \\
          Dequeue (\begin{tabular}[t]{@{}l@{}}
            non-empty case fixed, 
            empty case future)
          \end{tabular}
        \end{tabular}
        \\\mystrut
        Array-based queue 
        & \cite{CG05}${}^{\refeqn{degrees-1}}$
        &
        \begin{tabular}[t]{@{}l@{}}
          Enqueue (\begin{tabular}[t]{@{}l@{}}
            non-full case fixed, 
            full case future),      
          \end{tabular}
          \\
          Dequeue (\begin{tabular}[t]{@{}l@{}}
            non-empty case fixed,
            empty case future)
          \end{tabular}
        \end{tabular}
        \\\mystrut
        Lock coupling list
        & \cite{HS08}    & Add (fixed), Remove (fixed)
        \\
        \mystrut
        Lazy set  & \cite{HHLMSS07} & 
        \begin{tabular}[t]{@{}l@{}}
          Add (fixed), Remove (fixed), \\
          Contains (external)
        \end{tabular}
        \\\mystrut
        Elimination stack & \cite{HSY10} & Push (external), Pop (external)
        \\\mystrut
        HW queue${}^{\refeqn{degrees-2}}$ & \cite{Herlihy90}
        & Enqueue (future), Dequeue (future)
        \\\mystrut
        RDCSS 
        &  \cite{HFP02} & Restricted double-compare single-swap (future) 
        \\\mystrut
        CCAS &  \cite{FH07} & Conditional CAS (future) 
        \\\mystrut
        Elimination  queue &  \cite{MNSS05} & Enqueue (future),
        Dequeue (future) 
        \\\mystrut
        Snark-double ended queue
        & \cite{Doherty04DCAS}
        & 
        \begin{tabular}[t]{@{}l@{}}
          PushRight (future), PopRight (future), \\
          PushLeft (future), PopLeft (future)
        \end{tabular}
        \\\mystrut
        HM lock-free set 
        & \cite{Mic02}${}^{\refeqn{degrees-3}}$
        & \begin{tabular}[t]{@{}l@{}}
          Add (true case fixed, false case fut.), 
          \\
          Remove (true case fixed, false case fut.), 
          \\
          Contains (external)
        \end{tabular}
        \\\mystrut
        TSR Multiset & \cite{TSR14}
        & 
        \begin{tabular}[t]{@{}l@{}}
          Insert (fixed), 
          Delete (future), \\
          Lookup (external)
        \end{tabular}
      \end{tabular}
    \begin{footnotesize}
      \begin{enumerate}
      \item[]
      \item\label{degrees-1} This is a corrected version of the queue
        by Shann \etal \cite{SHC00}.
      \item\label{degrees-2} The dequeue operation is partial and retries
        as long as the queue is empty.
      \item\label{degrees-3} Algorithm is based on \cite{Har01}. 
      \end{enumerate}
    \end{footnotesize}
\caption{Classification of algorithms}
    \label{tab:degrees}
\rule{\textwidth}{1pt}
  \end{table}

Algorithms such as the Treiber stack have fixed (or static)
linearization points, whose execution only linearizes the operation to
which the linearization point belongs. These linearization points can
be conditional on the global state. For example, in the {\tt pop}
operation of the Treiber stack, the statement labelled {\tt P2} is a
linearization point for the empty case if {\tt Head = null} holds when
{\tt P2} is executed --- at this point, if \mbox{{\tt Head = null}}
holds, one can be guaranteed that the {\tt pop} operation will return
{\tt empty} and in addition that the corresponding abstract stack is
empty.  Proving correctness of such algorithms is relatively
straightforward, because reasoning may be performed in a forward
manner. In particular, for each atomic statement of the operation, one
can predetermine whether or not the statement is a linearization point
and generate proof obligations accordingly. In some cases, reasoning
can even be automated \cite{Vaf10}.

Unfortunately, not all algorithms can be verified in this manner. For
example, the {\tt contains} operation of the lazy set by Heller \etal
\cite{HHLMSS07} has external linearization points, where the operation
may potentially be linearized by the execution of a {\tt remove}
operation (by another concurrent process). The {\tt contains}
operation executing in isolation must set its own linearization
points, but interference from other processes may cause it to be
linearized externally.

A third, yet more complicated class of algorithms are those whose
linearization points depend on the future behaviour of the
algorithm. Reasoning here must be able to state properties of the
form: ``If in the future, the algorithm has some behaviour, then the
current statement is a linearization point.'' Further complications
arise when states of concrete system potentially corresponds to
several possible states of the abstract data type. Hence, for each
step of the concrete, one must check that each potential abstract data
type is modified appropriately. An example of such an algorithm is the
queue by Herlihy and Wing \cite{Herlihy90}, where each concrete state
corresponds to a set of abstract queues determined by the shared array
and the states of all operations operating on the array.  These
constitute the most difficult class of algorithms that have been
verified correct.

\begin{table}[!h]
  \rule{\textwidth}{1pt}
  \begin{tabular}{>{\bfseries}p{60mm}| 
      >{}p{60mm}
      >{\arraybackslash}p{20mm}}
        Method & \bfseries Algorithms verified & \bfseries Reference \\
        \hline\mystrut
        Canonical abstraction${}^{\refeqn{methods-fn-1}}$
        &         
        \begin{tabular}[t]{@{}l@{}}
          Treiber stack \\
          MS queue \\
          Array-based queue \\
          Lazy set \\
          Elimination stack \\
          Snark double-ended queue 
        \end{tabular}
        &      
        \begin{tabular}[t]{@{}l@{}}
          \cite{Groves09} \\
          \cite{DGLM04}${}^{\refeqn{methods-fn-2}}$\\
          \cite{CG05} \\
          \cite{CGLM06} \\
          \cite{GC07} \\
          \cite{Doherty04DCAS}
        \end{tabular}
        \\\mystrut
        Sequential abstraction
        &
        \begin{tabular}[t]{@{}l@{}}
          Treiber stack, Lock-coupling set \\
          Lazy set \\
          HW queue 
        \end{tabular}
        & 
        \begin{tabular}[t]{@{}l@{}}
          \cite{DSW11TOPLAS}\\
          \cite{DSW11}\\
          \cite{SDW14}
        \end{tabular}
        \\\mystrut
        RGSep (Rely-Guarantee separation logic)
        &
        \begin{tabular}[t]{@{}l@{}}
          RDCSS, Lock-coupling set, \\
          Optimistic set, 
          Lazy set \\
          MCAS \\ 
          CCAS, Elimination stack, \\
          Two-lock queue, 
          MS
          queue${}^{\refeqn{methods-fn-2}}$, \\ HM lock-free
          set${}^{\refeqn{methods-fn-4}}$
        \end{tabular}
        &
        \begin{tabular}[t]{@{}l@{}}
          \cite{Vaf07} and \\
          \cite{LF13-TR}\\
          \\
          \cite{Vaf07}\\ 
          \cite{LF13-TR}
        \end{tabular}
        \\\mystrut
        Reduction &
        \begin{tabular}[t]{@{}l@{}}
          Treiber stack  \\
          MS queue \\
          Elimination stack \\
          Simplified multiset 
        \end{tabular}
        & 
        \begin{tabular}[t]{@{}l@{}}
          \cite{EQSST10,Groves09}\\
          \cite{Gro08,EQSST10}\\
          \cite{GC09} \\
          \cite{EQSST10}
        \end{tabular}
        \\\mystrut
        RGITL (Rely-guarantee Interval Temporal Logic) &
        \begin{tabular}[t]{@{}l@{}}
          Treiber stack, 
          MS queue${}^{\refeqn{methods-fn-2}}$ \\
          Treiber stack
          with hazard pointers \\
          TSR multiset         
        \end{tabular}
        &
        \begin{tabular}[t]{@{}l@{}}
          \cite{BSTR11}\\
          \cite{TSR11} \\
          \cite{TSR14}
        \end{tabular}
        \\\mystrut
        Shape analysis &
        \begin{tabular}[t]{@{}l@{}}
          Treiber stack, MS queue${}^{\refeqn{methods-fn-2}}$  \\
          Numerous algorithms from \\
          \cite{HS08} 
        \end{tabular}
        & 
        \begin{tabular}[t]{@{}l@{}}
          \cite{ARRSY07}\\ 
          \cite{Vaf10}\\
          {}
        \end{tabular}
        \\\mystrut
        Construction-based &
        \begin{tabular}[t]{@{}l@{}}
          Treiber stack \\
          MS queue\\
          \\ 
          Elimination stack \\
          Optimistic set 
        \end{tabular}
        & 
        \begin{tabular}[t]{@{}l@{}}
          \cite{Jon12}\\ 
          \cite{Abrial05} \\
          and \cite{GC09}
          \\ 
          \cite{GC07} \\
          \cite{VY08}
        \end{tabular}
        \\\mystrut
        Hindsight lemma &
        \begin{tabular}[t]{@{}l@{}}
          Optimistic
          set${}^{\refeqn{methods-fn-3}}$, Lazy
          set${}^{\refeqn{methods-fn-3}}$
        \end{tabular}
        & 
        \begin{tabular}[t]{@{}l@{}}
          \cite{OHea10,ORVYY10TR}
        \end{tabular}
        \\\mystrut
        Interval abstraction & Lazy set & \cite{DD13AVoCS}          
        \\\mystrut
        Aspect-oriented proofs\label{} & HW queue & \cite{HSV13}
      \end{tabular}
    \begin{footnotesize}
      \begin{enumerate}
      \item []
      \item\label{methods-fn-1} This is
        the only method known to have found two bugs in existing
        algorithms  \cite{Doherty03,CG05}. 
      \item\label{methods-fn-2}Including
        a variation in \cite{DGLM04}. 
      \item\label{methods-fn-3} The use of atomicity brackets
        prohibits behaviours that are permitted by the
        fine-grained algorithm.
      \item \label{methods-fn-4} Set algorithm in \cite{Mic02}, which 
        is based on \cite{Har01}.
     \end{enumerate}
    \end{footnotesize}

      \caption{Methods for verifying linearizability}
    \label{tab:methods}
\rule{\textwidth}{1pt}
\end{table}

\reftab{tab:methods} presents a summary of methods for verifying
linearizability, together with the algorithms that have been verified
with each method and references to the papers in which the
verifications are explained. \reftab{tab:categories} then presents
further details of each method, the first column details whether or
not algorithms with fixed and external linearization points have been
proved, and the second details whether or not algorithms with future
linearization points have been proved. The third column details the
associated tool (if one exists), the fourth details whether the method
uses a compositional approach, and the fifth details whether each
method is known to be complete. The final column details whether the
methods have been linked formally to Herlihy and Wing's definitions of
linearizability.

\begin{table}[t]
  \rule{\textwidth}{1pt}
      \begin{tabular}[t]{>{\bfseries}m{34mm}| 
                         >{\centering}m{13mm} 
                         >{\centering}m{8mm} 
                         >{\centering}m{7mm}
                         >{\centering}m{16mm}
                         >{\centering}m{17mm}
                         >{\centering\arraybackslash}m{15mm}}
        Method & \textbf{Fixed/ External} & \textbf{Future} &
        \textbf{Tool} & \textbf{Compo\-sitional?} &
        \textbf{Complete?} & \textbf{Linked to HW}
        \\
        \hline \mystrut
        Canonical abstraction
        & $\tick$ & $\tick$ & PVS & & \refeqn{category-fn-1} &
        \refeqn{category-fn-4}
        \\\mystrut
        Sequential abstraction
        & $\tick$ & $\tick$ & KIV & & \refeqn{category-fn-2} & $\tick$
        \\\mystrut
        RGSep
        & $\tick$ & $\tick$ &  & $\tick$ & \refeqn{category-fn-3} & \refeqn{category-fn-5}
        \\\mystrut
        Reduction  & $\tick$ &  & QED & &  &  
        \\\mystrut
        RGITL & $\tick$  & $\tick$ & KIV & $\tick$ & \refeqn{category-fn-2} & \refeqn{category-fn-6}
        \\\mystrut
        Shape analysis  & $\tick$ &  & CAVE & &  &  
        \\\mystrut
        Construction-based & $\tick$ &  &  & &  &  
        \\\mystrut
        Hindsight lemma & $\tick$ &  &  & &  &   
        \\\mystrut
        Interval abstraction & $\tick$ & & & $\tick$ & & 
        \\\mystrut
        Aspect-oriented proofs\label{} &  & $\tick$ & CAVE & & & \refeqn{category-fn-7}
      \end{tabular}
    \begin{footnotesize}
      \begin{enumerate}
      \item []
      \item\label{category-fn-1} Forwards and backwards simulation is
        complete for showing refinement of input/output automata
        \cite{LynchV95}.
      \item\label{category-fn-2} Backward simulation for
        history-enhanced data types shown to be complete for
        linearizability Schellhorn \etal \cite{SWD12,SDW14}.
      \item\label{category-fn-3} Via completeness of auxiliary and
        prophecy variables for proving refinement
        \cite{AL91-existence}.
      \item\label{category-fn-4} Using results of Lynch
        \cite{Lyn96-DA}.
      \item\label{category-fn-5} Using results in \cite{LF13,LF13-TR}.
      \item\label{category-fn-6} Using an alternative characterisation
        of linearizability based on \emph{possibilities}
        \cite{Herlihy90}.
      \item\label{category-fn-7} Applies purely blocking
        implementations only.
      \end{enumerate}
    \end{footnotesize}

    \caption{Comparison of verification methods}
    \label{tab:categories}
    \vspace{-2mm}
\rule{\textwidth}{1pt}
  \end{table}

\subsection{Simulation-based verification}
\label{sec:simul-with-canon}

The first formal proofs of linearizability
\cite{CGLM06,CG05,CDG05,DGLM04,Doherty03} use simulation in the
framework of Input/Output Automata \cite{Lynch89}. Verification
proceeds with respect to \emph{canonical constructions}
\cite{Lyn96-DA}, where each operation call consists of an invocation,
a single atomic transition that performs the abstract operation, and a
return transition. The operations of a canonical object may be
interleaved meaning its histories are concurrent, but the main
transition is performed in a single atomic step. Lynch \cite{Lyn96-DA}
has shown that the history of every canonical construction is
linearizable, and hence, any implementation that refines it must also
be linearizable.

\begin{figure}[h]
  \rule{\textwidth}{1pt}\medskip

  \centering
  \scalebox{0.85}{\input{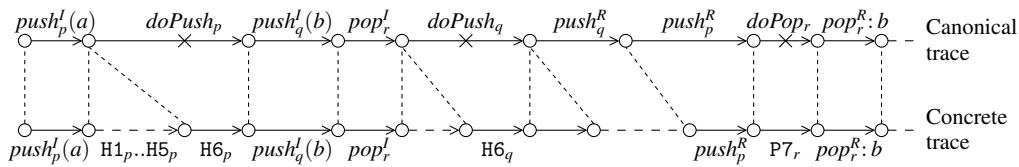}}
  \caption{Groves \etal's simulation proofs for linearizability}
  \label{fig:groves-sim}
\rule{\textwidth}{1pt}
\end{figure}

To demonstrate this technique, consider the trace from
\reffig{fig:sim}, recalling that the successful CAS statements at {\tt
  H6} and {\tt P7} are linearization points for the $push$ and
non-empty $pop$ operations, respectively. After proving simulation,
one obtains the mapping between the concrete and canonical traces
shown in \reffig{fig:groves-sim}. Namely, each invocation (response)
transition of the concrete maps to an invocation (response) of the
abstract, while a linearizing transition maps to a main
transition. The other concrete transitions are stuttering steps (see
\reffig{fig:sim}), and hence, have no effect on the canonical state.


Although Groves et al. present a sound method for proving
linearizability, a fundamental question about the link between
concurrent and sequential programs remains. \emph{ Can linearizability
  be formulated as an instance of data refinement between a concurrent
  implementation and a sequential abstract program?
}  
This is answered by Derrick \etal \cite{DSW11TOPLAS}, who present a
simulation-based method for proving refinement between a concurrent
and sequential (as opposed to canonical) object. Their methods include
an auxiliary history variable in the states of both the concrete and
abstract objects so that linearizability is established as part of the
refinement. In addition, a number of process-local proof obligations
that dispense with histories are generated, whose satisfaction implies
linearizability. Instead of proving refinement in a layered manner,
Derrick et al's proofs aim to capture the relationships between the
abstract and concrete systems within the refinement relation itself.

\begin{figure}[h]
  \rule{\textwidth}{1pt}\smallskip
  
  \centering
  \scalebox{0.85}{\input{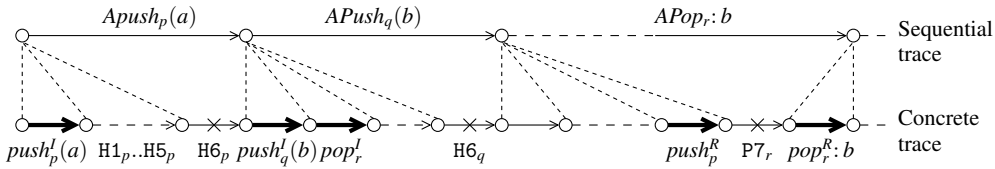}}
  \caption{Derrick \etal's refinement proofs for linearizability}
  \label{fig:derrick-sim}

\rule{\textwidth}{1pt}
\end{figure}

For a concrete example, once again consider the stack trace from
\reffig{fig:sim}.  Using the methods of Derrick \etal
\cite{DSW11TOPLAS}, one would obtain a refinement shown in
\reffig{fig:derrick-sim}, where the concrete transitions that update
the history are indicated with a bold arrow. Assume $hc$ and $ha$ are
the concrete and abstract history variables, both of which are
sequences of events. Each concrete invoke or return transition appends
the corresponding event to the end of $hc$, e.g., transition
$push_p^I(a)$ updates $hc$ to $hc \cat \aang{push_p^I(a)}$. Every
abstract transition updates the $ha$ with matching invocation and
response pairs, e.g., $APush_p$ updates the $ha$ to $ha \cat
\aang{push_p^I(a), push_p^R}$. Therefore, the concrete history $hc$
may be concurrent, whereas the abstract history $ha$ is
sequential. This enables the proof of linearizability to be built into
the refinement relation, as opposed to relying on a canonical
formulation to generate linearizable traces.

\subsection{Augmented states}
\label{sec:flatt-with-auxil}

Instead of defining concrete and abstract objects as separate systems
and using a representation relation to link their behaviours (as done
in \refsec{sec:simul-with-canon}), one may embed the abstract system
directly within the concrete system as an auxiliary extension
\cite{Vaf07} and prove linearizability by reasoning about this
combined system. For example, in the Treiber stack, one would
introduce an abstract sequence, say $Stack$, to the program in
\refsec{fig:TS}. At each linearization point of the Treiber stack, a
corresponding operation is performed on $Stack$, e.g., the successful
{\tt CAS} transition at {\tt H6} is augmented so that $Stack$ is
updated to $\aang{v} \cat Stack$
\cite{Vaf07}. 
This has the advantage of flattening the state space into a single
layer meaning proofs of linearizability follow from invariants on the
combined state. Vafeiadis \cite{Vaf07} further simplifies proofs by
using a framework that combines \emph{separation logic} \cite{ORY01}
(to simplify reasoning about pointers) and \emph{rely-guarantee}
\cite{Jon83} (to support compositionality). It is worth noting,
however, the underlying theory using this method relies on refinement
to prove linearizability \cite{LF13}. Namely, the augmentation of each
concrete state must be an abstraction of the concrete object.




\begin{figure}[h]
\rule{\textwidth}{1pt}

  \centering
  \scalebox{0.85}{\input{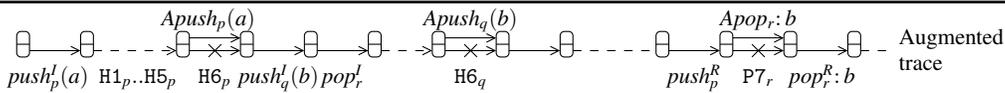}}
  \caption{Vafeiadis \etal's augmented state based proofs}
  \label{fig:vaf-sim}
\rule{\textwidth}{1pt}
\end{figure}

To visualise this approach, again consider the example trace from
\reffig{fig:interleaved}, where embedding the abstract state as an
auxiliary variable produces the augmented trace in
\reffig{fig:vaf-sim}.  For algorithms with fixed linearization points
(which can be verified using forward simulation), reasoning about
invariants over the flattened state space is simpler than simulation
proofs. (This is also observed in the forward simulation proof of
Colvin \etal \cite{CGLM06}, where auxiliary variables that encode the
abstract state are introduced at the concrete level.) However,
invariant-based proofs only allow reasoning about a single state at a
time, and hence are less flexible than refinement relations, which
relate a concrete state to potentially many abstract
states. 
Vafeiadis \cite{Vaf07} addresses these shortcomings using more
sophisticated auxiliary statements that are able to linearize both the
currently executing operation as well as other executing processes. In
addition, \emph{prophecy variables} \cite{AL91-existence} are used to
reason about operations whose linearization points depend on future
behaviour. Recently, Liang and Feng \cite{LF13} have consolidated
these ideas augmentations by allowing auxiliary statements {\tt
  linself}, (which performs the same function as the augmentations of
Vafeiadis by linearizing the currently executing process
\cite{VHHS06}) and {\tt lin(p)}, (which performs the linearization of
process $p$ different from $self$ that may be executing a different
operation). Liang and Feng (unlike Vafeiadis) allow augmentations that
use {\tt try} and {\tt commit} pairs, where the {\tt try} is used to
guess potential linearization points, and the {\tt commit} used to
pick from the linearization points that have been guessed thus far.

Augmented state spaces also form the basis for \emph{shape analysis}
\cite{JM82}, which is a static analysis technique for verifying
properties of objects with dynamically allocated
memory. 
One of the first shape-analysis-based linearizability proofs is that
of Amit \etal \cite{ARRSY07}, who consider implementations using
singly linked lists and fixed linearization points. The following
paraphrases \cite[pg 480]{ARRSY07}, by clarifying their nomenclature
with the terminology used in this paper.
\begin{quote}
  The proof method uses a correlating semantics, which simultaneously
  manipulates two memory states: a so-called \emph{candidate state}
  [i.e., concrete state] and the \emph{reference state} [i.e.,
  abstract state]. The candidate state is manipulated according to an
  interleaved execution and whenever a process reaches a linearization
  point in a given procedure, the correlating semantics invokes the
  same procedure with the same arguments on the reference state. The
  interleaved execution is not allowed to proceed until the execution
  over the reference state terminates. The reference response [i.e.,
  return value] is saved, and compared to the response of the
  corresponding candidate operation when it terminates. Thus
  linearizability of an interleaved execution is verified by
  constructing a (serial) witness execution for every interleaved
  execution.
\end{quote}
These methods are extended by Vafeiadis \cite{Vaf09-SVA}, where a distinction
is made between shape abstraction (describing the structure of a
concurrent object) and value abstraction (describing the values
contained within the object). The method is used to verify several
algorithms, including the complex RDCSS algorithm with external
linearization points.

Although the behaviours of concurrent objects are complex, many
algorithms that implement them are short, consisting of only a few
lines of code. This makes it feasible to perform a brute force search
for their linearization points. To this end, Vafeiadis presents a
fully automated method that considers all linearization points in a
single transition \cite{Vaf10}, and infers the required abstraction
mappings based on the given program and abstract specification of the
objects. The method thus far is only able to handle so-called
\emph{logically pure} algorithms, i.e., those that do not logically
modify the corresponding abstract state.

\subsection{Interval-based methods}
\label{sec:rely-guar-interv}

Interval-based methods aim to treat programs as executing over an
interval of time, as opposed to treating their statements as
transitions from a pre to post state.
Schellhorn \etal\ combine rely-guarantee reasoning with interval
temporal logic \cite{Mos00}, which enables one to reason over the
interval of time in which a program executes, as opposed to single
state transition \cite{STER11}. The proofs are carried out using the
KIV theorem prover \cite{DRSSSW93}, which is combined with symbolic
execution \cite{Bur74,BBNRS10} to enable guarantee conditions to be
checked by inductively stepping through the program statements within
KIV, simplifying mechanised verification. These methods have been
applied to verify the Treiber stack and the Michael/Scott queue
\cite{BSTR11}. 

Our own methods \cite{DD13AVoCS} verify behaviour refinement between a
coarse-grained abstraction and fine-grained implementation. The basic
motivation resembles the reduction-based approaches and hence, unlike
\cite{BSTR11,STER11,BBNRS10}, these methods simplify linearizability
proofs by allowing abstraction to be proved without having to identify
linearization points in the concrete code. This has been applied to
the lazy set algorithm \cite{HHLMSS07}.



\subsection{Problem-specific techniques.}
\label{sec:probl-spec-techn}
Concurrency researchers have also developed problem-specific methods,
sacrificing generality in favour of simpler linearizability proofs for
a specific subset of concurrent objects. One such method for
non-blocking algorithms is the Hindsight Lemma \cite{OHea10}, which
applies to linked list implementations of concurrent sets (e.g., the
lazy set) and characterises conditions under which a node is
guaranteed to have been in or out of a set. The original paper
\cite{OHea10} only considers a simple optimistic set. The extended
technical report \cite{ORVYY10TR} presents a proof of the Heller
\etal's lazy set. Unfortunately, the locks within the add and remove
operations are modelled using atomicity brackets, which has the
unwanted side effect of disallowing concurrent reads of the locked
nodes. That is, although O'Hearn \etal\ claim to have a proof of the
lazy set, their use of atomicity brackets, mean that the algorithm
they have verified is in fact \emph{not} the lazy set. Overall, the
ideas behind problem-specific simplifications such as the Hindsight
Lemma are interesting, but the logic used and the objects considered
are highly
specialised. 


Some objects like queues and stacks can be uniquely identified by
their aspects (properties that ensure the object in question has been
implemented). This is exploited by Henzinger \etal \cite{HSV13}, who
show present an aspect-oriented proof of the Herlihy/Wing
queue. Further details of this particular method are provided in
\refsec{sec:method-2:-aspect}.

Automation has been achieved for algorithms with helping mechanisms
and external linearization points such as the elimination stack
\cite{DGH13}. These techniques require the algorithms to satisfy
so-called $R$-linearizability \cite{SP93}, a stronger condition than
linearizability, hence, verification of algorithms with linearization
points based on future behaviour are excluded.

\subsection{Construction-based proofs}
\label{sec:line-constr}
Several researchers have also proposed development of linearizable
algorithms via incremental refinement, starting with an abstract
specification. Due to the transitivity of refinement, and because the
operations of the initial program are atomic, linearizability of the
final program is also guaranteed. An advantage of this approach is the
ability to \emph{design} an implementation algorithm, leaving open the
possibility of developing variations of the desired algorithm.

The first constructive approach to linearizability is by Abrial
\cite{Abrial05}, who use the Event-B framework \cite{Abrial2010-book}
and the associated proof tool. However, the final algorithm they
obtain requires counters on the nodes (as opposed to pointers
\cite{Michael96}), and it is not clear whether such a scheme really is
implementable.  Groves \cite{Gro08-Deriv} presents a derivation of the
Michael/Scott queue using reduction to justify each refinement step
\cite{Lip75}.  This is extended by Groves and Colvin \cite{GC09}, a
more complicated stack by Hendler \cite{HSY10} is derived. This stack
uses an additional backoff array in the presence of high contention
for the shared central stack. Their derivation methods allow data
refinement (without changing atomicity), operation refinement (where
atomicity is modified, but state spaces remain the same) and
refactoring (where the structure of the program is modified without
changing its logical meaning) \cite{GC09,GC07}. These proofs are not
mechanised, but there is potential to perform mechanisation using
proof tools such as the QED \cite{Elmas10}.

Gao et al.~\cite{GFH09,GH07,GGH07,GGH05} present a number of
derivations of non-blocking algorithms, including via the use of
special-purpose reduction theorems \cite{GH07}. However, these
derivations aim to preserve {\em lock-freedom} (a progress property)
\cite{MP92-lf}, as opposed to linearizability.

Vechev et al.~\cite{VY08,VYY09} present tool-assisted derivation
methods based using bounded model checking to obtain assurances that a
derived algorithm is linearizable. Starting with a sequential
linked-list set, they derive variations of the set algorithm
implemented using DCAS and CAS instructions, as well as variations
that use marking schemes. Although their methods allow relatively
large state spaces to be searched, these state spaces are bounded in
size, and hence, only finite executions are checked; linearizability
verification requires potentially infinite executions to be verified.

More recently, Jonsson \cite{Jon12} has presented a derivation of the
Treiber stack and Michael/Scott queue in a refinement calculus
framework \cite{MV92}. Jonsson defines linearizability using: \emph{ A
  program $P$ is linearizable if and only if $atomic \{P\}$ is refined
  by $P$ \cite[Definition 3.1]{Jon12}.} Reduction-style commutativity
checks are used to justify splitting the atomicity at each stage. With
such an interpretation of linearizability, one is able able to start
by treating the entire concrete operation as a single atomic
transition, then incrementally split its atomicity into finer-grained
portions. 

\section{Case study 1: An optimistic set algorithm}
\label{sec:case-study:-fine}

\begin{table}[t]
  \rule{\textwidth}{1pt}    
      \begin{tabular}[t]{>{\bfseries}p{20mm}| 
                         >{\centering}p{40mm}
                         >{\arraybackslash}p{60mm}}
                       \begin{tabular}[t]{@{}l@{}}
                         Reference
                       \end{tabular}
                       &
        \bfseries 
        Lin. point identification
        &
        \begin{tabular}[t]{@{}l@{}}
          \bfseries Additional notes 
        \end{tabular}\\
        \hline\mystrut
        \cite{VHHS06} & Manual & 
        Operation {\tt
          contains} not verified 
        \\
        \mystrut
        \cite{CGLM06} & Manual & 
        Allows model checking 
        \\        \mystrut
        \cite{Vaf07} & Manual  &
        Auxiliary code can 
        linearize other operations
        \\        \mystrut
        \cite{Vaf10}
        & Automatic & Full automation via shape analysis, but the lazy set \cite{HHLMSS07} is not yet
          verified in the method.
        \\        \mystrut
        \cite{OHea10} & N/A & 
        Uses
        Hindsight Lemma to generate proof obligations, and hence, only applicable to
        list-based set implementations
        \\\mystrut
        \cite{EQT09}&
        N/A  
        &
        Linearizability proofs are
        performed for coarse-grained abstractions
        \\\mystrut
        \cite{DSW11} & Manual 
        &
        Data refinement-based proofs
        \\\mystrut
        \cite{LF13-TR} & Manual 
        & 
        Separation logic encoding
        \\\mystrut
        \cite{DD13AVoCS} & N/A 
        &  
        Interval-based reasoning; linearizability is proved for
        coarse-grained abstractions 
      \end{tabular}
    \caption{Overview of methods for verifying set algorithms}
    \label{tab:overview}
\rule{\textwidth}{1pt}  \end{table}

Set algorithms have become standard case studies for showing
applicability of a theory to verifying linearizability.  Of particular
interest is the lazy set by Heller \etal \cite{HHLMSS07}, which is a
simple algorithm with \Add and \Rem operations that have fixed
linearization points and a \Cont operation that is potentially
linearized by the execution of other operations. We first present a
verification of a simplified version that consists of \Add and \Rem
operations only.
An overview of the different approaches to verifying set algorithms is
given in \reftab{tab:overview}. Further details of each method are
provided in the sections that follow. 
The formalisation in this section aims to highlight the main ideas
behind each method. We refer readers interested in reproducing each
proof to the original papers.


  

\OMIT{

An early proof attempt of the lazy set using auxiliary variables is
that of Vafeiadis \etal\ \cite{VHHS06}, who extend the concrete
state space with additional variables corresponding to the abstraction
of the object in question, then augment linearization points
of the concrete code with statements corresponding to the abstract
operation.  The methods in \cite{VHHS06} are unable to verify the {\tt
  contains} operation. This is addressed in Vafeiadis' thesis
\cite{Vaf07} where augmentations are able to reason about the
linearization points of other system processes, enabling a proof of
the lazy set algorithm.

The first complete proof is that of Colvin \etal \cite{CGLM06}, who
show that the concrete implementation simulates an abstract canonical
algorithm. Their proofs use the framework of Input/Output Automata
\cite{Lynch89} and are mechanised in the PVS theorem prover
\cite{owre96pvs}. Their setup allows invariants to be checked in a
model checker, providing some level of assurance that the final proof
will be successful. Verification of \Cont requires the use of
backwards simulation \cite{deRoever98} because the proof must refer to
future behaviour of the operation.

Vafeiadis presents a tool-supported method that allows linearization
points to be automatically inferred \cite{Vaf10}. Although a proof of
the lazy set is not given (due to difficulties in automatically
inferring correct abstractions), the method is able to verify other
set algorithms that requires reference to linearization points outside
the operation. The clear advantage of this method is that when an
algorithm is within the scope of CAVE \cite{Vaf10}, no input
additional input from the verifier is needed.

O'Hearn \etal\ present a methodology specific to sets
that are implemented concretely by linked lists is given by
\cite{OHea10,ORVYY10TR}, however, simplifications made about the
atomicity of locked sections of code mean that (contrary to the claims
in the paper), the algorithms verified are in fact \emph{not} the
optimistic nor lazy sets. In particular, by making locked sections of
code atomic, the algorithms in \cite{OHea10,ORVYY10TR} do not allow
traversal of the locked sections while the actual addition/removal is
being performed.

Elmas \etal\ present a method based on \emph{reduction} \cite{Lip75}
that enables incremental replacement of fine-grained code in an
implementation by statements with coarser-grained atomicity. Their
methods use the QED verifier \cite{Elmas10} and are used to prove a
multiset algorithm consisting of \Add and \Rem operations, but
without a \Cont operation \cite{EQT09}. Reduction is aimed at
simplifying linearizability proofs and reducing the complexity of an
implementation, and therefore do not need to identify the
linearization points in the concrete code. Note that the multiset
algorithm is also used as a case study by Travkin \etal\ \cite{TWS12},
but here, the only operations considered are \Add and
\Cont. Linearizability could not be proved with the inclusion of a
{\tt delete} operation. Using the framework of Rely-Guarantee
Interval Temporal Logic \cite{STER11}, which allows compositional
reasoning. An advantage of their framework is that it enables symbolic
execution of a algorithm and its properties within the KIV theorem
prover \cite{DRSSSW93}.

Derrick \etal\ return to refinement-based proofs \cite{DSW11},
presenting a verification of the lazy set.  \emph{Non-atomic
  refinement} \cite{DW05} and \emph{potential linearization points}
are encode future behaviours of the \Cont operation and the
proof have been mechanised in KIV. Their techniques build on
\cite{DSW11TOPLAS} and use history enhanced data types, and hence, are
the first to link data refinement of implementations to Herlihy and
Wing's original definition of linearizability \cite{Herlihy90}.

Liang and Feng \cite{LF13,LF13-TR} present an approach that extends
the original ideas of Vafeiadis \etal\ \cite{VHHS06}, where
linearization points of the algorithm in question are augmented with
auxiliary statements that describe behaviour at the linearization
points. 

The final approach we consider is our own method, which uses a
framework based on interval predicates \cite{DD13AVoCS}. Here, we
consider the behaviour of the algorithm in question over its interval
of execution, then show that this behaviour implies the behaviour of a
coarse-grained algorithm over the same interval. Like the
reduction-based methods, our approach is aimed at simplifying proofs
of linearizability by transforming the fine-grained algorithm to
another algorithm with coarser-grained atomicity, and hence, requires
a second proof that the coarse-grained algorithm is itself
linearizable.  }

\OMIT{ The optimistic set algorithm is presented in
  \refsec{sec:simple-set}, its linearization points are discussed in
  \refsec{sec:verify-fixed-line}, and three different proofs of
  linearizability are presented in Sections
  \ref{sec:simul-based-proofs}, \ref{sec:refin-based-proofs}, and
  \ref{sec:auxiliary-variables}. }

\subsection{An optimistic set}
\label{sec:simple-set}
In this section, we present a simplified version of Heller \etal's
concurrent set algorithm \cite{HHLMSS07} (see \reffig{fig:lazyset-ar})
operating on a shared linked list, which is sorted in strictly
ascending values order. Locks are used to control concurrent access to
list nodes. The algorithm consists of operations \Add and {\tt remove}
that use auxiliary operation {\tt locate} to optimistically determine
the position of the node to be inserted/deleted from the linked
list. \begin{figure}
\rule{\textwidth}{1pt}  \centering
  \footnotesize
    
    \begin{sidebyside}[3]
      \tt add(x):
    
      \ A1: n1, n3 := locate(x);

      \ A2: \textbf{if} n3.val != x \textbf{then}

      \ A3: \ \ n2 := \textbf{new} Node(x);

      \ A4: \ \ n2.next := n3;

      \ A5: \ \ n1.next := n2;

      \ A6: \ \ res := true

      \ \ \ \ \ \textbf{else}

      \ A7: \ \ res := false


      \ A8: n1.unlock();

      \ A9: n3.unlock();

      A10: \textbf{return} res
      
      \nextside

      \tt remove(x):
    
      \ R1: n1, n2 := locate(x);

      \ R2: \textbf{if} n2.val = x \textbf{then}

      \ R3: \ \ n2.mark := true;

      \ R4: \ \ n3 := n2.next;

      \ R5: \ \ n1.next := n3;

      \ R6: \ \ res := true

      \ \ \ \ \ \textbf{else}

      \ R7:\ \ \ res := false;


      \ R8:\ n1.unlock();

      \ R9:\ n2.unlock();

      R10:\ \textbf{return} res

      \nextside 

    \tt
    locate(x):

    \ \ \ \ \ \textbf{while} true \textbf{do}
    
    \ L1: \ \ pred := Head; 

    \ L2: \ \ curr := pred.next; 

    \ L3: \ \ \textbf{while} curr.val < x \textbf{do} 

    \ L4: \ \ \ \ pred := curr; 

    \ L5: \ \ \ \ curr := pred.next
    

    \ L6: \ \ pred.lock();
    
    \ L7: \ \ curr.lock();

    \ L8: \ \ \textbf{if} !pred.mark 

    \ \ \ \ \ \ \ \ \ \textbf{and} !curr.mark 
    
    \ \ \ \ \ \ \ \ \ \textbf{and} pred.next = curr
    
    \ L9:\ \ \ \textbf{then} return pred, curr

    \ \ \ \ \ \ \ \textbf{else} 

    L10:\ \ \ \ pred.unlock(); 
    
    L11:\ \ \ \ curr.unlock()
    
  \end{sidebyside}
    


    

    
  \caption{Optimistic set algorithm  operations}
  \label{fig:lazyset-ar}
\rule{\textwidth}{1pt}\end{figure}

Each node of the list consists of fields $val, next, mark$, and
$lock$, where $val$ stores the value of the node, $next$ is a pointer
to the next node in the list, $mark$ denotes the marked
bit\footnote{The $mark$ bit is not strictly necessary to implement the
  optimistic set (e.g., \cite{Vaf07}), however, we use it here to
  simplify the lead up to the lazy set in \refsec{sec:case-study-2}. } and $lock$ stores
the identifier of the process that currently holds the lock to the
node (if any). The $lock$ field of each node only prevents
modification to the node; it is possible for processes executing
{\tt locate} and \Cont to read values of locked nodes
when they traverse the list.  Two dummy nodes with values $-\infty$
and $\infty$ are used at the start ({\tt Head}) and end ({\tt
  Tail}) of the list. All values $v$ inserted are assumed to satisfy
$-\infty < v < \infty$.

Operation {\tt locate(x)} is used to obtain pointers to two nodes {\tt
  pred} (the predecessor node) and {\tt curr} (the current node). A
call to {\tt locate(x)} operation traverses the list ignoring locks,
acquires locks once a node with value greater than or equal to {\tt x}
is reached, then \emph{validates} the locked nodes. If the validation
fails, the locks are released and the search for {\tt x} is
restarted. When {\tt locate(x)} returns, both {\tt pred} and {\tt
  curr} are locked by the calling process, the value of {\tt pred} is
always less than {\tt x}, and the value of {\tt curr} may either be
greater than {\tt x} (if {\tt x} is not in the list) or equal to {\tt
  x} (if {\tt x} is in the list).


Operation {\tt add(x)} calls {\tt locate(x)}, then if {\tt x}
is not already in the list (i.e., value of the current node {\tt n3}
is strictly greater than {\tt x}), a new node {\tt n2} with value
field {\tt x} is inserted into the list between {\tt n1} and {\tt n3}
and {\tt true} is returned. If {\tt x} is already in the list, the
{\tt add(x)} operation does nothing and returns {\tt false}. Operation
{\tt remove(x)} also starts by calling {\tt locate(x)}, then if {\tt
  x} is in the list the current node {\tt n2} is removed and {\tt
  true} is returned to indicate that {\tt x} was found and removed. If
{\tt x} is not in the list, the \Rem operation does nothing and
returns {\tt false}. Note that operation {\tt remove(x)} distinguishes
between a logical removal, which sets the marked field of {\tt n2}
(the node corresponding to {\tt x}), and a physical removal, which
updates the {\tt next} field of {\tt n1} so that {\tt n2} is no longer
reachable.

\begin{figure}[h]
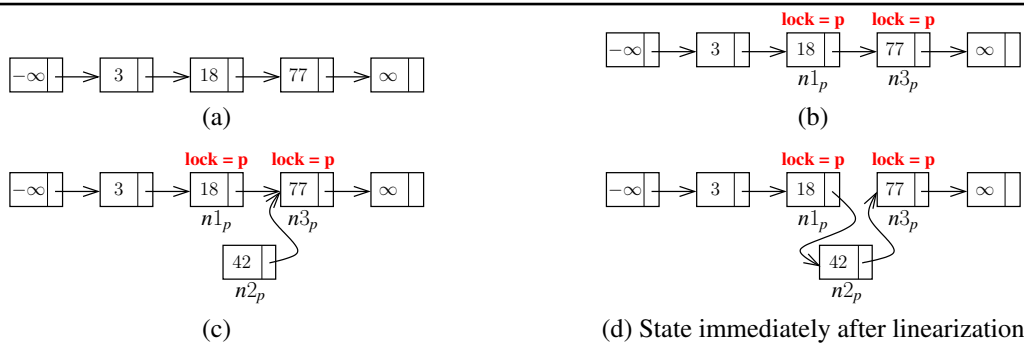

\rule{\textwidth}{1pt}  \centering
  \begin{minipage}[t]{0.475\columnwidth}
    \centering
    \scalebox{0.5}{\input{add1.pspdftex}}\\
    (a) 
    \\[2mm]
    \scalebox{0.5}{\input{add5.pspdftex}}
    \\
    (c)
  \end{minipage}
  \begin{minipage}[t]{0.475\columnwidth}
    \centering
    \scalebox{0.5}{\input{add4.pspdftex}}
    \\
    (b)
    \\[2mm]
    \scalebox{0.5}{\input{add6.pspdftex}}\\
    (d) State immediately after linearization
  \end{minipage}
  \caption{Execution of {\tt add(42)} by process $p$}
  \label{fig:add1}
\rule{\textwidth}{1pt}\end{figure}

As a concrete example, consider the linked list in \reffig{fig:add1}
(a), which represents the set $\{3,18,77\}$, and an execution {\tt
  add(42)} by process $p$ without interference. Execution starts by
calling {\tt locate(42)}, which searches for the predecessor ({\tt
  pred}) and successor ({\tt curr}) of the node to be added, ignoring
any other locks (lines {\tt L3}-{\tt L5}). After exiting the loop, the
executing process locks both {\tt pred} and {\tt curr}, which prevents
their modification by other processes (lines {\tt L6}-{\tt L7}). At
{\tt L8}, the process checks to ensure that both {\tt pred} and {\tt
  curr} are unmarked (which ensures that they have not been removed
since the end of the loop) and that {\tt pred.next} is still {\tt
  curr} (which ensures that no new nodes have been inserted between
{\tt pred} and {\tt curr} since the end of the loop). If the test at
{\tt L8} fails, the locks on {\tt pred} and {\tt curr} are released
and the process {\tt locate} is restarted. In our example execution,
we assume that the test is successful, which causes $n1_p$ and $n2_p$
to be set as shown in \reffig{fig:add1} (b). Having found and locked
the correct location for the insertion, the process executing \Add
tests to see that the value is not already in the set (line {\tt A2}),
then creates a new unmarked node $n3_p$ with value $42$ and next
pointer $n3_p$ (see \reffig{fig:add1} (c)). Then by executing {\tt
  A4}, the executing process sets the next pointer of $n1_p$ to $n2_p$
causing a successful \Add operation to be linearized (see
\reffig{fig:add1} (d)). Thus, provided no {\tt remove(42)} operations
are executed, any other {\tt add(42)} operation that is started after
{\tt A4} has been executed will return $false$.  After the
linearization, process $p$ releases the locks on $n1_p$ and $n3_p$ and
returns $true$ to indicate the operation was successful.

\begin{figure}[h]
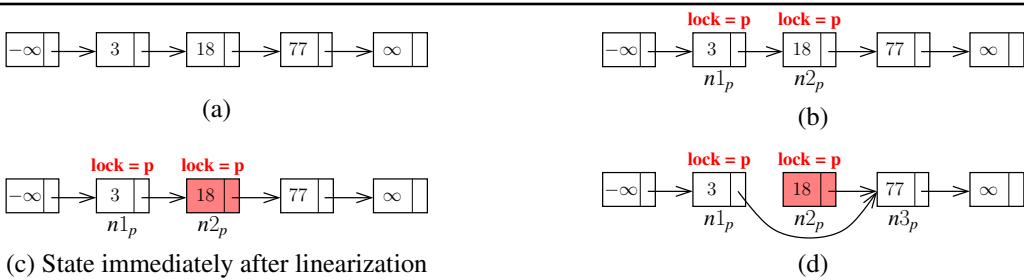

\rule{\textwidth}{1pt}  \centering
  \begin{minipage}[b]{0.475\columnwidth}
    \centering
    \scalebox{0.5}{\input{rem.pspdftex}}\\[2.2mm]
    (a)
    \\[4mm]
    \scalebox{0.5}{\input{rem3.pspdftex}}
    \\
    (c) State immediately after linearization
  \end{minipage}
  \begin{minipage}[b]{0.475\columnwidth}
    \centering
    \scalebox{0.5}{\input{rem2.pspdftex}}\\
    (b)\\[2mm]
    \scalebox{0.5}{\input{rem4.pspdftex}}\\
    (d)
  \end{minipage}
  \caption{Execution of {\tt remove(18)} by process $p$}
  \label{fig:rem}
\rule{\textwidth}{1pt}\end{figure}

Now consider the execution of {\tt remove(18)} by process $p$ on the
set $\{3,18,77\}$ depicted by the linked list in \reffig{fig:rem} (a),
where the process executes without interference. Like {\tt add}, the
\Rem operation calls {\tt locate(18)}, which returns the state
depicted in \reffig{fig:rem} (b). At {\tt R2}, a check is made that
the element to be removed (given by node $n2_p$) is actually in the
set. Then, the node $n2_p$ is removed logically by setting its marked
value to $true$ (line {\tt R3}), which is the linearization point of
\Rem (see \reffig{fig:rem} (c)).  After execution of the linearization
point, operation \Rem sets $n3_p$ to be the next pointer of the
removed node (line {\tt R4}), and then node $n2_p$ is physically
removed by setting the next pointer of {\tt n1} to $n3_p$ (see
\reffig{fig:rem} (d)). Then, the held locks are released and $true$ is
returned to indicate that the \Rem operation was been successful. Note
that although $18$ has been logically removed from the set in
\reffig{fig:rem} (c), no other process is able to insert $18$ to the
set until the marked node has also been physically removed (as
depicted in \reffig{fig:rem} (d)), and the lock on $n1_p$ has been
released.

\paragraph{Verifying \Add and \Rem operations}

Verifying correctness of \Add and {\tt remove}, which have fixed
linearization points is relatively straightforward because the
globally visible effect of both operations may be determined without
having to refer to the \emph{future states} of the linked list. The
refinement-based methods (Section
\ref{sec:simul-with-canon}) 
verify correctness using forward simulation and the state augmentation
methods (\refsec{sec:flatt-with-auxil}) modify the abstract state
directly. 

We present outlines of the proofs using the simulation-based methods
of Colvin \etal \cite{CGLM06} (\refsec{sec:simul-based-proofs}),
refinement-based method of Derrick \etal \cite{DSW11}
(\refsec{sec:refin-based-proofs}) and auxiliary variable method of
Vafeiadis \cite{Vaf07}
(\refsec{sec:auxiliary-variables}).  
To unify the presentation, we translate the PVS formulae from
\cite{CGLM06-PVS} and the Vafeiadis' RGSep notation \cite{VP07,Vaf07}
into Z \cite{z:bowe96}, which is the notation used by Derrick \etal\
Inevitably, this causes some of the benefits of a proof method to be
lost; we discuss the effect of the translation and the benefits
provided by the original framework, where necessary.

Full details on modelling concurrent algorithms with Z are given in
\cite{DSW11TOPLAS}. To reason about linked lists, memory must be
explicitly modelled, and hence, the concrete state $CState$ is defined
as follows, where $Label$ and $Node$ are assumed to be the types of a
program counter label and node, respectively. Each atomic program
statement is represented by a Z schema.  For example, the schema for
the statements in \reffig{fig:lazyset-ar} labelled {\tt A5} and {\tt
  A7} executed by process $p$ are modelled by $Add5_p$ and $Add7_p$,
respectively. Notation $\Delta CState$ imports both unprimed and
primed version of the variables of $CState$ into the specification
enabling one to identify specifications that modify $CState$; unprimed
and primed variables are evaluated in the current and next states,
respectively. Using the Object-Z \cite{Smith99} convention, we assume
that variables $v' = v$ for every variable $v$ unless $v' = k$ is
explicitly defined for some value $k$.

\begin{small}
  \begin{sidebyside}[3]
    \begin{schema}{CState}
      pred,curr : P \fun Node\\
      n1,n2,n3 : P \fun Node\\
      pc : P \fun Label \\
      lock : Node \fun \power P\\
      next : Node \fun Node\\
      mark : Node \fun \bool\\
      res : P \fun V
    \end{schema}
    \nextside
    \begin{schema}{Add5_p}
      \Delta CState \ST
      pc(p) = A5 \\
      next'(n1(p)) = n2(p)
    \end{schema}
    \nextside
    \begin{schema}{Add7_p}
      \Delta CState \ST
      pc(p) = A7 \\
      res'(p) = false
    \end{schema}
  \end{sidebyside}
\end{small}

\subsection{Method 1: Proofs against canonical specifications}
\label{sec:simul-based-proofs} 

Following \cite{DGLM04}, one is required to perform the following
steps.\vspace{-1.5mm}
\begin{enumerate} 
\item Identify and fix the linearization points of each concrete
  operation. 
\item Define a canonical abstraction and a representation relation
  that describes the link between the canonical and concrete
  representations.
\item Prove simulation between the concrete program (which is the
  program in \reffig{fig:lazyset-ar} formalised in Z) and canonical
  abstraction, where the concrete initialisation and responses are
  matched with abstract initialisation and response operations,
  respectively. The linearization points must be matched with main
  canonical operations. Simulation may be performed in a forwards or
  backwards manner, and in some cases, both are required. Furthermore,
  the proof may require introduction of additional invariants at the
  concrete level to specify additional properties of the data
  structure in question.

\end{enumerate}\vspace{-1.5mm}

The linearization points have been described in
\refsec{sec:simple-set}. To model the canonical specification, first
the abstract state $AState$ must be defined.
\begin{sidebyside}
  \begin{schema}{AState}
    S : \power V \\
    pc : P \fun Label\\
    v : P \fun V\\
    res : P \fun \bool
  \end{schema}
\end{sidebyside}
The canonical operations corresponding to the \Add operation are given
by the following Z schema, where variables decorated with $?$ and $!$
denote inputs and outputs, respectively.
\begin{small}
  \begin{sidebyside}[4]
    \begin{schema}{AddInv_p}
      \Delta AState \\
      x? \in V
      \ST
      pc(p) = idle \\
      pc'(p) = addi \\
      v'(p) = x?
    \end{schema}
    \nextside
    \begin{schema}{AddOK_p}
      \Delta AState \ST
      pc(p) = addi \\
      v(p) \in S \\
      S' = S \cup \{v(p)\} \\
      res'(p) = true \\
      pc'(p) = addo
    \end{schema}
    \nextside
    \begin{schema}{AddFail_p}
      \Delta AState \ST
      pc(p) = addi \\
      v(p) \notin S \\
      res'(p) = false \\
      pc'(p) = addo
    \end{schema}
    \nextside
    \begin{schema}{AddRes_p}
      \Delta AState \\
      r! \in \bool\ST
      pc(p) = addo \\
      r! = res(p) \\
      pc'(p) = idle \\
    \end{schema}
  \end{sidebyside}
\end{small}

Similar schema are generated for the canonical form of the \Rem
operation. Following Lynch \cite{Lyn96-DA}, any history generated by
such canonical specifications are linearizable, and therefore, any
refinement of the canonical specification must also be linearizable.

As highlighted in \refsec{fig:sim}, the forward simulation must
consider four different simulation diagrams: initialisation,
stuttering and non-stuttering transitions, and finalisation. For the
non-stuttering transitions, (which are the most interesting of these)
the forward simulation proof rule states the following, where $AOp_p$
is the abstract operation corresponding to the $COp_p$ in process $p$,
$rep$ is a relation from the abstract to the concrete state space, and
`$\semi$' denotes relational composition, i.e., for relations $r_1 \in
V_X \leftrightarrow V_Y$ and $r_2 \in V_Y \leftrightarrow V_Z$, we
define $r_1 \semi r_2 = \{(x, z) | x \in V_X \land z \in V_Z \land
\exists y : V_Y \st (x, y) \in r_1 \land (y, z) \in
r_2\}$. 
\begin{eqnarray}
  \label{eq:11}
   \all p:  P  \st rep \semi COp_{p} \subseteq AOp_{p} \semi rep
\end{eqnarray}
Thus, for any abstract state $\sigma$ and concrete state $\tau$ linked
by the representation relation $rep$, if the concrete statement
$COp_p$ is able to transition from $\tau$ to $\tau'$, then there must
exist an abstract state $\sigma'$ such that $AOp_p$ can transition
from $\sigma$ to $\sigma'$ and $\sigma'$ is related to $\tau'$ via
$rep$.

Colvin \etal \cite{CGLM06} set up a framework that enables model
checking of possible invariants prior to its formal verification in a
theorem prover. To this end, auxiliary variables that reflect the
abstract space are introduced at the concrete level together with
invariants over these auxiliary variables that correspond to the
simulation relation. For the lazy set, one such variable is $aux\_S$,
which stores the set of elements currently in the set. The set
$aux\_S$ is updated whenever a node is inserted into the list, or is
marked for deletion. To verify that $aux\_S$ does indeed represent the
abstract set, one must prove that the following holds: \vspace{1mm}

\hfill
$cs(aux\_S)  =  \{k \in V | InList(cs,k)\}
$\hfill{} \vspace{1mm}

\noindent where $cs$ is a reachable concrete state and $InList$ is a
function that determines whether or not the value $k$ is in the list
(i.e., an unmarked node with value $k$ is reachable from the head).
The main invariants that Colvin et al. \cite{CGLM06,CGLM06-PVS} prove
are:\footnote{In \cite{CGLM06-PVS} $A5$ and $A7$ are labelled $add6$
  and $add8$, respectively.}
\begin{eqnarray}
\all p : P \st pc(p) \in \{A5, R7\} 
\imp v(p) \notin aux\_S
\label{eq:4}
\\
\all p : P \st pc(p) \in \{A7, R3\} 
\imp v(p) \in aux\_S
\label{eq:5}
\end{eqnarray}
By \refeq{eq:4}, for any process $p$, prior to execution of
execution of {\tt A5} (a successful {\tt add}) and {\tt R7} (a
failed {\tt remove}), the element being added and removed,
respectively must not be in the set. Condition \refeq{eq:5} is
similar.  The representation relation between an abstract state $as$
and concrete state $cs$ is defined as follows, where $step\_rel$ is a
relation between the program counters of $as$ and $cs$. \vspace{1mm}

\hfill
$rep(as, cs)   \sdef as(S) = cs(aux\_S)
  \land step\_rel(as, cs)
$\hfill{} \vspace{1mm}

\noindent Proofs of these conditions require a number of additional
invariants to be established, e.g., stating that the list is
sorted. 
However, it is worth noting that a substantial number of these
invariants are introduced to prove the full lazy set. These proofs are
carried out entirely within PVS \cite{owre96pvs}.

\subsection{Method 2: Proofs against sequential specifications}
\label{sec:refin-based-proofs}

Derrick \etal's method consider proofs directly against a sequential
specification. 
Verification using this
method consists of the following steps.\vspace{-1.5mm}
\begin{enumerate} 
\item Identify and fix the linearization points of each
  operation. 
\item Prove that individual concrete runs correctly implement the
  abstract operations, i.e., that every sequence of fine-grained
  transitions correctly produces the corresponding abstract
  coarse-grained transition.
\item Show that other processes running in parallel maintain the
  refinement relation. To this end, encode the {\em interference
    freedom} and {\em disjointness} proof obligations within the
  invariants.
\item Decompose the proof into process-local proof obligations using a
  \emph{status} function.
\item Finally, guarantee correct initialisation.
\end{enumerate}\vspace{-1.5mm}

The abstract state and operations \Add and \Rem are modelled as
follows:
\begin{eqnarray*}
  AState & \sdef &  [ S : \power V] 
  \\
  Add_p & \sdef & [\Delta AState, x? : V, 
  r! : \bool |
  S' = S \union \{x?\}\land 
  r! = (S' \neq S)] \\
  Remove_p & \sdef & [      \Delta AState,
  x? : V,
  r! : \bool |
  S' = S \setminus \{x?\}\land
  r! = (S' \neq S) ]
\end{eqnarray*}
The proofs rely on \emph{history-enhanced objects}, which introduce
the sequential and concrete histories as auxiliary
variables. Executing operations append events to a history, e.g., an
invocation $op$ with input $x$ executed by process $p$, appends
$inv(p,op,x)$ to the history. The abstract data types execute all
operations atomically, and hence, their invocation and return occur as
part of a single transition. Given that $hs$ is the auxiliary
sequential histories, the following formalises the history-enhanced
add and remove operations:
  \begin{eqnarray*}
    AddH_{p} & \sdef &
    \begin{array}[t]{@{}l@{}}
      Add_{p} \land 
      {[}hs,hs' : \seq(Event) | hs' = hs
      \cat \lseq inv(p,add,x?), ret(p,add,r!)\rseq {]}
    \end{array}
    \\
    RemH_{p} & \sdef &
    \begin{array}[t]{@{}l@{}}
      Remove_{p} \land 
      {[}hs,hs' : \seq(Event) | hs' = hs
      \cat \lseq inv(p,rem,x?), ret(p,rem,r!)\rseq ]
    \end{array}
  \end{eqnarray*}
Similarly, if $h$ is the concrete history variable, the invocation and
return schema of the \Add operation are extended as follows:
\begin{eqnarray*}
  AddInvH_{p} & \sdef &  AddInv_{p} \land [h,h' : \seq(Event) | h' = h
  \cat \lseq inv(p,add,x?)\rseq ]\\
  AddRetH_{p} & \sdef &  AddRet_{p} \land [h,h' : \seq(Event) | h' = h
  \cat \lseq ret(p,add,r!)\rseq ]
\end{eqnarray*}
Therefore, the abstract history is sequential, whereas the concrete is
concurrent. Refinement between the abstract and concrete
history-enhanced data types must explicitly prove linearizability
between the two histories. 


The proofs here involve showing that each process is a {\em non-atomic
  refinement} \cite{DeWe03,DW05} of the abstract data type. To relate
the concrete history $h$ to an abstract history $hs$, Derrick \etal\
use an additional set $R$ that stores a set of return events for
pending invocations whose effects have taken place, and therefore
contributes to $hs$. In particular, assuming $\bseq(X)$ denotes
bijective sequences of type $X$, some $h_0 \in \bseq(R)$ can be used
as the $h_0$ that completes pending invocations. The set $\bseq(R)$
contains all sequences constructed from $R$, so that each element of
$R$ appears in the sequence exactly once. Each process $p$ may or may
not contribute a return event $ret(p,i,out)$ to the set $R$. If it
does, then there must be a pending invocation $inv(p,i,x)$ in concrete
history $h$, and $p$ must have already passed the linearization point
and therefore modified the representation of the abstract object to
implement $AOp_{p}$. If it does not, then it either does not execute
an operation at all, or it has a pending invocation $inv(p,i,x)$ in
its history.

The proof obligations refer to the (of type $STATUS \abssynt IDLE \mid
IN \lang V \rang \mid OUT \lang V \rang$) of each process. Namely,
process $p$ has status $IDLE$ iff $p$ is not executing any operation,
$IN(x)$ iff $p$ is executing an operation with input $x$, but has not
passed the linearization point of the operation, and $OUT(r)$ iff $p$
is executing an operation and has passed the linearization point with
return value $r$.  This is combined with a function $runs : CState
\times P \fun O \cup \{none\}$ denoting the operation the given
process is executing in a given state ($none$ if the process is idle)
and a function $status: CState \times P \rightarrow STATUS$, which
determines whether or not the process contributes a return event in a
given state. The encoding of the $status$ is such that is $IDLE$ if
$runs(cs, p) = none$; is $IN(x)$ if $runs(cs, p) = op$ and $cs(pc(p))$
is not past the linearization point of $op$; and is $OUT(r)$ if
$runs(cs, p) = op$ and $cs(pc(p))$ is past the linearization point of
$op$.

The forward simulation relation $rep$ is then of the following form,
where $pi(n, h)$ denotes that $h(n)$ is a pending invocation event,
i.e., $h(n)$ is an invocation and for all $m > n$, $h(m)$ is not a
return event that matches $h(n)$.
\begin{align}
  & rep((as, hs),(cs, h)) \defs \nonumber \\
  \label{eq:1}
  &\qquad ABS(as,cs) \land INV(cs) \land (\all p, q \st p \neq q \imp D(cs, p, q))\\
  \label{eq:8}
  &\qquad  \land  (\forall n \st pi(n,h) \implies runs(cs,
  proc(h(n))) = oper(h(n))) \\
  \label{eq:9}
  &\qquad  \land 
    \begin{array}[c]{@{}l@{}}
      \forall p, x \st \begin{array}[t]{@{}l@{}}
        status(cs, p) = IN(x) 
        \implies \exists n \st pi(n,h) \land h(n) = inv(p,runs(cs, p),x)
      \end{array}
    \end{array}
    \\
  \label{eq:10}
  &\qquad \land \exists R \st
  \begin{array}[t]{@{}l@{}}
    R = \{ret(p,op,r) | runs(cs,p) =
    op
    \land     status(cs, p) = OUT(r)\} \\
    \land \forall h_0 : \bseq(R) \st  linearizable(h, h_0, hs)
  \end{array}
\end{align}
Here, \refeq{eq:1} states that both abstraction $ABS$ and invariant
$INV$ hold, and that $D(cs, p, q)$ holds, which ensures interference
freedom for the local states of process $p$ are not modified by
execution of process $q$. Conjunct \refeq{eq:8} states that if $h(n)$
is a pending invocation, then function $runs$ is accurate. Conjunct
\refeq{eq:9} states that whenever process $p$'s status is $IN(x)$ for
some $x$, there must exist an index $n \in \dom(h)$ such that $h(n)$
is a pending invocation, and corresponds to an invocation that is
executing $runs(cs,p)$ with input $x$. Finally conjunct \refeq{eq:10}
relates $h$ to $hs$ using the set of processes with status $OUT$. It
requires that there exist a set $R$ of events corresponding to
processes that have executed a linearizing statement, but not yet
returned, such that for any bijective sequence $h_0$ generated from
$R$, $linearizable(h, h_0, hs)$ holds.

Finally, a number of process-local\footnote{Thread-local in the
  terminology of Derrick \etal \cite{DSW11TOPLAS}.} proof obligations
that do not need to refer to histories $hs$ and $h$ are generated, and
a theorem that ensures satisfaction of the process-local properties
that implies $rep$ holds. These proof obligations use information from
the $status$ function to determine the correct simulation
condition. For example, the proof obligation below is for steps of
process $p$ that transition from a status $IN(in)$, where $COp_p$
potentially corresponds to the execution of a linearization point.
$$\begin{array}[t]{@{}l@{}}
  \forall as:AS, cs,cs' : CState, p: P \st
  rep(as, cs) \wedge status(cs,p) = IN(in)
  \wedge COp_p(cs, cs')  \implies \\
  \qquad \begin{array}[t]{@{}l@{}}
    status(cs',p) = IN(in) \wedge rep(as, cs') {} \vee {}\\ 
    (\exists as', out \dot AOp_p(in, as, as', out)
    \wedge status(cs',p) = OUT(out)  \wedge rep(as', cs'))
  \end{array}  
\end{array}$$

\noindent 
Verifying invocation and response transitions are straightforward
because the abstraction is not modified, and stuttering transitions
are straightforward because the histories are not modified. The
non-stuttering transitions linearize the abstract object. This is
reflected in the $status$ function, whose value changes from $IN(x)$
before the transition to $OUT(r)$ after the transition. Locality of
the proof method is guaranteed using the well-established technique of
non-atomic refinement \cite{DW05} (we refer the interested reader to
\cite{DSW11TOPLAS} for details).

\subsection{Method 3: Augmented states and RGSep}
\label{sec:auxiliary-variables}

The method of Vafeiadis \cite{Vaf07}, requires the following
steps. \vspace{-1.5mm}
\begin{enumerate}
\item Introduce auxiliary variables to the existing program, at least
  one of which is an abstraction of the data type in question. Define
  the abstract operations on these auxiliary variables that are
  required to be implemented by the concrete program.
\item Identify the linearization points of the concrete
  implementation, then introduce the appropriate auxiliary statements
  at each linearization point.
\item Define a rely condition by identifying statements that modify
  the global state, and developing an abstraction of each
  statement. The overall rely is a disjunction of each such
  abstraction.
\item Define and prove an invariant that links the abstract and
  concrete representations. 
\end{enumerate}\vspace{-1.5mm}
Vafeiadis' proofs are performed using the RGSep framework
\cite{Vaf07,VP07}. In this paper, for uniformity, we translate the
example expressed in RGSep into
Z. 

For the \Add operation, a state space is extended with a fresh
variables {\tt AbsRes} (representing the abstract result) and {\tt S}
(representing the abstract set) to obtain an augmented state
$AugState$. In addition, the fixed linearization points {\tt A5} and
{\tt A7} are augmented as follows, where the brackets {\tt < >} denote
that {\tt stmt} within the brackets is executed atomically.
\vspace{1mm}

\hfill 
  \begin{minipage}{.95\columnwidth}
    \tt \small add(x):
  
    \ ...
  
    \ A5: <n1.next := n2; AbsRes := (x $\notin$ S); S := (S $\cup$ \{x\})>
  
    \ ...

    \ A7: <res := false;  AbsRes := (x $\notin$ S); S := (S $\cup$
    \{x\})> ...
  \end{minipage}
\hfill {} \vspace{1mm}

\noindent Note that at {\tt A7}, the auxiliary code sets {\tt AbsRes}
to $false$ (i.e., {\tt x $\notin$ S}), and therefore the abstract set
{\tt S} remains unchanged. The \Rem operation is similar, therefore
its details are elided.

Translating Vafeiadis' separation logic notation into Z, and
simplifying the notational overhead, we obtain the following
relations. Function $lock(n)$ returns the id of the process that
currently holds the lock on node $n$, where $lock(n) = \emptyset$
holds if no process has locked $n$. Assuming that $val(n)$, $next(n)$
and $mark(n)$, denote the value, next and mark fields of $n$, we
define shorthand: \vspace{1mm}

\hfill
$\begin{array}[t]{@{}l@{}}
lvn(n)  \sdef  (lock(n), val(n), next(n)) \qquad 
lvnm(n) \sdef  (lock(n), val(n), next(n), mark(n)) 
\end{array}$\hfill {} \vspace{1mm}

\noindent 
The non-stuttering actions of a program's environment are abstracted
by rely conditions, which are relations on the pre-post states
representing transitions that modify the global
state\footnote{\cite{Vaf07} defines provisos for some of these
  actions, which suggests that they be interpreted as implication. For
  consistency with the Z formalism, we formalise these proviso
  predicates as preconditions of each action.}. Because the abstract
and concrete state spaces are disjoint, we replace all instances of
separating conjunction `$*$' by logical conjunction `$\land$', which
enables simpler comparison among the different methods. We discuss the
differences that arise from this translation where
needed. \vspace{1mm}

\noindent
\begin{small}
  $\begin{array}[t]{@{}r@{~}c@{~}l@{}}
    AugState & \sdef & CState \land [S : \power V] \\
    Lock_p & \sdef & {[}\Delta AugState, n : Node | lock(n) =
    \emptyset \land p \neq 0 \land
    lock'(n) = \{p\} {]}\\
    Unlock_p & \sdef & {[} \Delta AugState, n : Node | lock(n) = \{p\}
    \land p \neq 0 \land 
    lock'(n) = \emptyset] 
    \\
    Mark_p & \sdef & \left[ \Delta AugState, n, n1 : Node, v : V
      \begin{array}[c]{@{~}|@{~}l@{}}
        lvn(n1) = (p, v, n) \land
        lvnm'(n1) = (p, v, n, true) \land \\
        S' = S \backslash \{v\}
      \end{array}
    \right]
    \\
    Add_p & \sdef & \left[
      \begin{array}[c]{@{}l@{}}
        \Delta AugState, \\
        n1, n2, n3 : Node,
        u, v: Val
      \end{array}
      \begin{array}[c]{@{~}|@{~}l@{}}
        (u < v < val(n3)) \land lvn(n1) = (p, u, n3) \land \\
        lock(n3) = \{p\} \land lvn(n2) = (\emptyset, v, n3) \land \\
        next'(n1) = n2 \land S' = S \cup \{v\}
      \end{array}
    \right]
    \\
    Remove_p & \sdef & \left[
      \begin{array}[c]{@{}l@{}}
        \Delta AugState , \\
        n1,n2,n3 : Node,
        u, v : V
      \end{array}
      \begin{array}[c]{@{~}|@{~}l@{}}
        lvn(n1) = (p, u, n2) \land
        lvnm(n2) = (p, v, n3, true) 
        \land\\
        lvn'(n1) = (p, u, n3)
      \end{array}
    \right]
  \end{array}$
\end{small}
\hfill{} \vspace{1mm}

The rely condition for process $p$ is \vspace{1mm}

\hfill $Rely_p  \sdef  \bigvee_{q \in P \bs \{p\}} Lock_q \lor Unlock_q \lor Add_q \lor
  Mark_q \lor Remove_q$ \hfill {} \vspace{1mm}

\noindent 
which describes the potential global modifications that the
environment of process $p$ can make.  With this encoding, one can
clearly see that the rely condition is an abstraction of statements of
\Add and \Rem that modify the global state.

\setlength{\fboxsep}{.3mm} Vafeiadis \cite{Vaf07} requires annotation
of code using separation logic-style assertions. In addition, building
on the framework of \cite{Jon83}, these assertions must be stable with
respect to the rely conditions. The proof outlines for the lazy set
are elided in \cite{Vaf07}, however, may be reconstructed from the
other list examples in the thesis. We further adapt the proof outlines
using Z-style notation. The invariants are formalised using the
following predicates, where $ls(x, A, y)$ converts the linked list
from $x$ to $y$ into an sequence $A$, predicate $sorted(A)$ holds iff
$A$ is sorted in ascending order, and $s(A)$ returns the set of
elements corresponding to $A$.
\begin{eqnarray*}
  ls(x, A, y) & \sdef &
  \begin{array}[t]{@{}l@{}}
    (x = y \land A = \aang{}) \lor \\ 
    \exists v,z,B \st x \neq y \land A = \aang{v} \cat B \land val(x)
    = v \land next(x) = z  \land ls(z, B, y)
  \end{array}
  \\
  sorted(A) & \sdef &
  \begin{array}[t]{@{}l@{}}
    \textbf{if}\ A \in \{\aang{~}, \aang{a}\}\
    \textbf{then}\ true\\ \textbf{elseif}\ A =
    \aang{a,b} \cat B\  \textbf{then}\ (a < b) \land sorted(\aang{b}
    \cat B)\ \textbf{else}\ false
  \end{array}
  \\
  s(A) & \sdef & S = \ran(A) \bs \{-\infty, \infty\}
\end{eqnarray*}

Note that due to a typographical error, the
failed case of the \Add operation is missing in \cite{Vaf07}, however,
it can be reconstructed from the \Rem operation (see
\reffig{fig:vaf-po}).
\begin{figure}[t]
\rule{\textwidth}{1pt}  \centering
  \begin{center}\small
    \begin{tabular}[t]{@{}l@{}}
      {\tt add(x) :}
      \\
      \ \ \begin{tabular}[t]{@{}l@{}}
        {\tt ...} 
        \\
        $\left\{
          \begin{array}[c]{@{}l@{}}
            \exists u,v \st
            \begin{array}[t]{@{}l@{}}
              \exists n,A, B \st  \begin{array}[t]{@{}l@{}}
                ls(Head, A, n1_p) \land lvn(n1_p) = (p, u, n3_p)  \\
                \land lvn(n3_p) = (p, v, n) \land ls(n, B, Tail) \land s(A \cat \aang{u, v} \cat B)
              \end{array}
              \\
              \land lvn(n2_p) = (\emptyset, x, n3_p) \land u < x \land x < v
            \end{array}
          \end{array}\right\}$
        \\
        \\[-1mm]
        \ {\tt A5: <n1.next := n2; AbsRes := (x $\notin$ S); S := (S $\cup$
          \{x\})>}
        \\[1mm]
        $\left\{
          \begin{array}[c]{@{}l@{}}
            \exists u, v \st  
            \exists n,A,B \st
            \begin{array}[t]{@{}l@{}}
              ls(Head, A, n1_p)
              \land lvn(n1_p) = (p, u,
              n2_p)  \\
              \land lvn(n2_p) = (p, x, n) \land ls(n, B, Tail) \land s(A\cat\aang{u,x} \cat B)
            \end{array}
          \end{array}
        \right\}$
        \\
        {\tt ...}
        \\
        $\left\{\exists n, A,B\st  ls(Head, A, n3_p) \land
          lvn(n3_p) = (p, x, n) \land 
          ls(n, B, Tail) \land s(A \cat \aang{x} \cat B)\right\}$
        \\[1mm]
        \ {\tt A7:   <res := false;  AbsRes := (x $\notin$ S); S := (S $\cup$ \{x\})>}
        \\[1mm]
        $\left\{\exists A \st  ls(Head, A, Tail) \land s(A)\right\}$
        \\
        {\tt ...}
      \end{tabular}
    \end{tabular}
  \end{center}
  
  \caption{Reconstructed proof outline for {\tt add(x)}}
\label{fig:vaf-po}
\rule{\textwidth}{1pt}\end{figure}
Of course, such annotations are not available in Z, but can easily be
encoded as invariants on the overall specification by explicitly
introducing a program counter variable. For example,
given that $pc(p)$ denotes the program counter for process $p$, whose
value is a program label, the assertion at ${\tt A7}$ can be encoded
as a predicate:\vspace{1mm}

\hfill
$\begin{array}[t]{rcl}
  POA7_p & \sdef & 
  pc(p)  = {\tt A7}  \imp 
  \exists n, A,B\st
  \begin{array}[t]{@{}l@{}}
    ls(Head, A, n3_p) \land lvn(n3_p) = (p, x, n) \land  
    \\ ls(n, B, Tail) \land 
    s(A \cat \aang{x} \cat B)
  \end{array}
\end{array}$\hfill{}\vspace{1mm}

\noindent Such proof obligations must be resilient to interference
from other processes \cite{OG76-Acta},
hence, one must verify that the following holds for each $p, q \in P$
such that $p \neq q$, where $Env_q \in \{Lock_q, Unlock_q, Mark_q,
Add_q, Remove_q\}$.  \vspace{1mm}

\hfill 
$POA7_p \semi Env_q \imp POA7_p'$
\hfill {} \vspace{1mm}


Liang and Feng \cite{LF13-TR} provide outlines for the \Rem and \Cont
operations albeit using a different framework, and define a number of
additional predicates prior to the proof for \Rem and {\tt
  contains}. These predicates largely mimic Vafeiadis' rely
conditions. As with Vafeiadis' proofs, a translation of Liang and
Feng's formalisation to Z is also possible. Due to the similarities
between the proof methods, we elide the details of such a
transformation in this paper.

\subsection{Discussion}
\label{sec:discussion}

With the advances in linearizability verification, correctness of the
optimistic set is straightforward, and there is even the possibility
of automating the verification procedure (e.g., by extending methods
in \cite{Vaf10,DGH13}\footnote{Note that the optimistic set in
  \cite{Vaf10} does not use a marking scheme, and hence, is different
  from the algorithm in \refsec{sec:simple-set}. The methods in
  \cite{DGH13} can only automatically verify algorithms with helping
  mechanisms.}). We have presented a detailed account of three methods
that manually identifies the linearization points, as well as
abstraction relations and invariants. These methods are based on
differing formal foundations: method 1 uses I/O Automata, method 2
uses Z, and method 3 uses RGSep. To simplify comparison between these
approaches, we have translated each of these to Z. An advantage of
RGSep (method 3) that is lost in the translation to Z is the ability
to syntactically distinguish between predicates that may be affected
by the environment. However, as already discussed, the majority of
predicates in each assertion are non-local, and hence, the loss of
this feature does not overly affect the proof. The proofs using
methods 1 and 2 are mechanised. Tool support for extensions to method
3 have been developed, and there is a possibility for mechanising
proofs using method 3 directly, but this has thus far not been
done. Each of the methods supports process-local verification. Method
1 proves invariants that describe the behaviours of the other
processes, method 2 explicitly encodes interference freedom conditions
in the refinement relation, and method 3 additionally supports
compositionality via rely-guarantee reasoning.

The underlying challenges in verifying linearizability are manifested
in each of the proof methods in essentially the same way. Namely, the
identification of the correct abstraction relations and invariants,
correct identification of linearization points and the corresponding
abstract changes that occur at each linearization point. These also
remain the difficult aspects of a proof to automate.

\section{Case study 2: A lazy set algorithm}
\label{sec:case-study-2}

In this section, we present the full lazy set algorithm, which
consists of a \Cont operation in addition to the \Add and \Rem
operations from the optimistic set. 
\refsec{sec:lazy-set-algorithm} presents the \Cont operation in
detail. Despite the simplicity of the \Cont operation, its
verification introduces significant complexity in the proof methods,
requiring the use of more advanced verification techniques.  Sections
\ref{sec:method-1:-simulation} and \ref{sec:method-2:-refinement}
present simulation-based proof methods with respect to canonical and
sequential abstract specifications, respectively, and
\refsec{sec:method-3:-auxiliary} present a method based on augmented
states.

\subsection{The \Cont operation}
\label{sec:lazy-set-algorithm}

A process executing {\tt contains(x)} traverses the list (ignoring
locks) from $Head$, stopping as soon as a node with value greater or
equal to {\tt x} is found. Value {\tt true} is returned if the node is
unmarked and its value is equal to {\tt x}, otherwise {\tt false} is
returned. Unlike {\tt locate}, the \Cont operation does not acquire
locks, and performs at most one traversal of the linked list.

\begin{figure}
\rule{\textwidth}{1pt}  \centering
  \begin{minipage}[b]{.34\columnwidth}
    \small
    \tt 
    contains(x) : 
    
    C1: curr := Head; 

    C2: \textbf{while} curr.val < x \textbf{do}

    C3: \ \ curr := curr.next
    

    C4: res := !curr.mark \textbf{and} 

    \ \ \ \ \ \ \ \ \ \ \ (curr.val = x)
    
    C5: \textbf{return} res

    \caption{\rm The \Cont operation}
    \label{fig:lazyset}
  \end{minipage}
  \hfill
  \begin{minipage}[b]{.32\columnwidth}
    \small
    \tt contains(x) :
  
    \dots 

    \dots

    C4a: r1 :=  !curr.mark; 

    C4b: res := r1 \textbf{and} 
    
    \ \ \ \ \ \ \ (curr.val = x);
  
    \dots 

    {}
    
    {}
    \caption{\rm Splitting atomicity (a)}
    \label{fig:lazyset-a}
  \end{minipage}
  \hfill
  \begin{minipage}[b]{.32\columnwidth}
    \small
    \tt contains(x) :

    \dots 
  
    \dots 

    C4c: r1 := (curr.val = x);

    C4d: res :=  !curr.mark \textbf{and} 

    \ \ \ \ \ \ \ \ \ \ \ \ r1; 
  
    \dots 

    \caption{\rm Splitting atomicity (b)}
    \label{fig:lazyset-b}
  \end{minipage}
\rule{\textwidth}{1pt}\end{figure}

When verifying linearizability of the \Cont operation, atomicity
constraints of an implementation often dictate that the expression in
{\tt C4} to be split. However, because the order in which the
variables within a non-atomic expression are accessed is not known,
there are two possible evaluation orders: \reffig{fig:lazyset-a} and
\reffig{fig:lazyset-b}, both using a temporary variable {\tt r1}. To
verify linearizability of the original operation in
\reffig{fig:lazyset}, both orders of evaluation must be
verified. However, Derrick et al. and Vafeiadis only consider
\reffig{fig:lazyset-a}, while Groves et al. only consider
\reffig{fig:lazyset-b}. It is also possible to consider both
possibilities at the same time using logics that enable reasoning
about the non-determinism in expression evaluation under concurrency
\cite{HBDJ13}, which is the approach taken in \cite{DD13AVoCS}.

\begin{figure}[h]
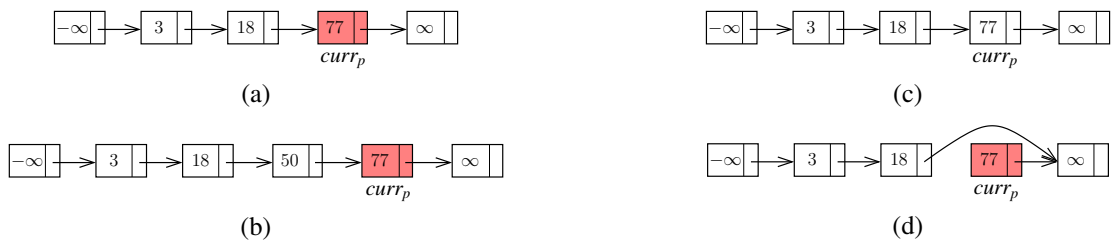

\rule{\textwidth}{1pt}\medskip

  \centering
  \begin{minipage}[t]{0.475\columnwidth}
    \centering
    \scalebox{0.48}{\input{cont1.pspdftex}}\\
    (a)
    \\[4.5mm]
    \scalebox{0.48}{\input{cont2.pspdftex}}
    \\
    (b)
  \end{minipage}
  \hfill 
  \begin{minipage}[t]{0.475\columnwidth}
    \centering
    \scalebox{0.48}{\input{cont3.pspdftex}}
    \\
    (c)\\[2mm]
    \scalebox{0.48}{\input{cont4.pspdftex}}
    \\
    (d)
  \end{minipage}
  \caption{Counter-examples for {\tt C4a} and {\tt C4b} as
    linearization points of \Cont}
  \label{fig:cont}
\rule{\textwidth}{1pt}\end{figure}

Unlike the \Add and \Rem operation, none of the statements of \Cont
qualify as valid linearization points. To see this, we consider the
two most suitable candidates, i.e., {\tt C4a} and {\tt C4b}, and
present counter-examples to show that neither of these are valid. The
essence of the issue is that a verifier must decide whether or not the
\Cont will return $true$ or $false$ (i.e., as its future behaviour) by
considering the state of the shared object when {\tt C4a} or {\tt C4b}
is executed, and this is impossible.  Suppose {\tt C4a} is chosen as
the linearization point of the \Cont operation. Now consider the
following state of the shared linked list in \reffig{fig:cont} (a),
where process $p$ is executing {\tt contains(50)} and has just exited
its loop because $curr_p.val \geq 50$, but has not yet executed
statement {\tt C4a}.  Suppose another process $q$ executes {\tt
  add(50)} to completion. This results in the linked list in
\reffig{fig:cont} (b), which corresponds to an abstract state
$\{3,18,50\}$.  Execution of process $p$ from this state will set
$r1_p$ to $false$, and hence the {\tt contains(50)} will return
$false$, even though the element $50$ is in the set (corresponding to
the shared linked list) when {\tt C4a} is executed.

Similarly, suppose {\tt C4b} is chosen to be the linearization point
of the \Cont operation. Assume there are no other concurrent
operations and that process $p$ is executing {\tt contains(77)} on the
linked list in \reffig{fig:rem} (a), and execution has reached (but
not yet executed) statement {\tt C4b}. This results in the state of
the linked list in \reffig{fig:cont} (c).  Suppose another process $q$
executes a {\tt remove(77)} operation to completion. This results in
\reffig{fig:cont} (d), corresponding to the abstract queue $\{3,18\}$.
Now, when process $p$ executes {\tt C4b}, it will set $res_p$ to
$true$, and hence, return $true$ even though $77$ is not in the
abstract set corresponding to the shared linked list when {\tt C4b} is
executed. Therefore, neither {\tt C4a} nor {\tt C4b} are appropriate
linearization points for \Cont.

Proving linearizability it turns out must consider the execution of
other operations, i.e., the linearization point cannot be determined
statically by examining the statements within the \Cont operation
alone. Here, \Cont may be linearized by the execution of an \Add or a
\Rem operation. As Colvin \etal\ point out: \emph{ The key to proving
  that [Heller et al's] lazy set is linearizable is to show that, for
  any failed {\tt contains(x)} operation, {\tt x} is absent from the
  set at some point during its execution \cite{CGLM06}.  } That is,
within any interval in which {\tt contains(x)} executes and returns
$true$, there is some point in the interval such that the abstract set
corresponding to the shared linked list contains $x$. Similarly, if
{\tt contains(x)} returns $false$, there is some point in the interval
of execution such that the corresponding abstract set does not contain
$x$. The statement that removes $x$ from the set is also responsible
for linearizing any {\tt contains(x)} operations that may return
false.

From a refinement perspective, the need for backward simulation arises
in executions in which the abstract specification resolves its
non-determinism earlier than the concrete implementation, resulting in
a future concrete transition that cannot be matched with an
abstract transition when the forward simulation rule \refeq{eq:11} is
used. 
Instead proofs must be performed using \emph{backward simulation}
\cite{deRoever98}, which for a non-stuttering transition generates a
proof obligation of the form: \vspace{1mm}

\hfill
$  \all p: P \st COp_p \semi rep \subseteq rep \semi AOp_p$\hfill {} \vspace{1mm}

\noindent 
This states that if $COp_p$ can transition from $\tau$ to $\tau'$ and
$\tau'$ is related by $rep$ to some abstract state $\sigma'$, then
there must exist an abstract state $\sigma$ such that $rep$ holds
between $\tau$ and $\sigma$ and $AOp_p$ can transition from $\sigma$
to $\sigma'$.  Such proofs involve reasoning from the end of
computation to the start, hence, are more complicated than forward
simulation. Equivalent to this is an encoding using prophecy variables
\cite{AL91-existence,Vaf07,ZFFSL12}.

\subsection{Method 1: Proofs against canonical automata}
\label{sec:method-1:-simulation}


Colvin \etal\ split their simulation proofs by introducing an
intermediate specification, that ``eliminates the need to know the
future \cite[pg 481]{CGLM06}''. They then prove backward simulation
between the canonical and intermediate specifications and forward
simulation between the intermediate and concrete specifications. To
simplify the backward simulation, the intermediate specification is
kept as similar to the canonical abstraction as possible.

The intermediate state introduces a local boolean variable
$seen\_out(p)$ that holds for a process $p$ executing {\tt
  contains(x)} iff {\tt x} has been absent from the abstract set since
$p$ invoked {\tt contains(x)}. The invocation of the intermediate {\tt
  contains(x)} operation sets $seen\_out(p)$ to $false$ if $x$ is in
$S$ and to $true$ otherwise. Furthermore, when the main transition of
a {\tt remove(x)} occurs, in addition to linearizing itself, it
linearizes all invoked {\tt contains(x)} operations that have not yet
set their $res(p)$ value.  Therefore, $IContInv_p$ (which invoke
the \texttt{contains}) and $IRemOK_p$ (which performs the main
remove operation) in the intermediate specification are defined as
follows: \vspace{1mm}

\hfill$\begin{array}[t]{rcl}
  IContInv_p(x) & \sdef & [\Delta IState, x? : V | ContInv_p(x?) \land seen\_out'(p) = (x? \notin S)]\\
  IRemOK_p(x) & \sdef &
  \left[\begin{array}[c]{@{}l@{}}
      \Delta IState, x? : V \begin{array}[c]{@{~~~}|@{~~~}l}
        RemOK_p(x?) \land \\
        \all q : P \st pc(q) = CIn \land v(q) = v(p) \imp \\
        \qquad\quad seen\_out'(q)
        = true
      \end{array}
    \end{array}
  \right]
\end{array}$\hfill {} \vspace{1mm}

\noindent 
The intermediate {\tt
  contains(x)} operation is allowed to return $false$ whenever
$seen\_out(x)$ holds, therefore, $ContFail_p$ is replaced by
$IContFail_p$ below:
\begin{eqnarray*}
  IContFail_p \sdef   [\Delta IState | pc(p) = CIn \land seen\_out(p)
  \land res'(p) = false \land pc'(p) = COut]\label{eq:12}
\end{eqnarray*}
Unlike $ContFail_p$, schema $IContFail_p$ can set 
$res(p)$ to $false$ even if $v(p) \in S$ holds in the
current state. This is allowed because whenever $pc(p) = CIn \land
seen\_out(p)$ holds, a state for which $v(p) \notin S$ holds must have
occurred at some point since the invocation of the \Cont
operation. When $pc(p) = CIn \land seen\_out(p) \land v(p) \in S$
holds, both $IContOK_p$ and $IContFail_p$ are enabled and process $p$
may non-deterministically choose to match with $res(p) = true$ (in the
current state) or with $res(p) = false$ (having linearized at some
point in the past).

The backward simulation relation $bsr$ below between the canonical and
intermediate state spaces is relatively straightforward because there
is no data refinement between intermediate state $is$ and abstract
state $as$. In particular, one obtains:\vspace{1mm}

\hfill
\begin{small}
  $
  \begin{array}[t]{rcl}\small
    bsr(is, as) & \sdef &
    \begin{array}[t]{@{}l@{}}
      is(S) = as(S) \land \\
      \all p \st 
      \begin{array}[t]{@{}l@{}}
        is(pc(p)) = as(pc(p)) \lor 
        \left(
          \begin{array}[c]{@{}l@{}}
            is(pc(p)) = CIn \land is(seen\_out(p)) \land \\
            as(pc(p)) = COut \land as(res(p)) = false
          \end{array}\right)
      \end{array}
    \end{array}
  \end{array}
  $
\end{small}
\hfill {} \vspace{1mm}

\noindent 
The second disjunct within the universal quantification is needed
because $p$ may have already executed $ContFail_p$ in the abstract,
and decided to return $false$, whereas the corresponding intermediate
operation has not yet made its choice. This delay in the intermediate
specification is only allowed if $seen\_out(p)$ holds in the
intermediate state.

A forward simulation is then used to prove refinement between the
intermediate and concrete systems. As in
\refsec{sec:simul-based-proofs}, this proof is simplified by
introducing an auxiliary set $aux\_S$ to the concrete code, which is
updated in the same way as in \refsec{sec:simul-based-proofs}. The
proof allows the same forward simulation to be used, but additional
invariants related to the \Cont operation must be introduced. For
example:
\begin{eqnarray}
  \all p \st
  \left(\begin{array}[c]{@{}l@{}}
      (pc(p) = 4c \land val(curr(p)) \neq v(p))  
      \lor \\
      (pc(p) \in \{4c, 4d\}  \land mark(curr(p))) 
    \end{array}\right)
  & \imp &  seen\_out(p)
  \label{eq:2}
  \\
  \all p \st pc(p) = 4d \land \neg mark(curr(p)) & \imp & v(p) \in aux\_S
  \label{eq:3}
\end{eqnarray}
By \refeq{eq:2}, if the concrete program is in a position to return
$false$, it must have already seen that the value being searched is not
in the set, and by \refeq{eq:3}, if the concrete program is in a
position to return true, the value being searched must be in the set. 
The proof of forward simulation then proceeds in a standard manner. 


\subsection{Method 2: Proofs against sequential specifications}
\label{sec:method-2:-refinement}

Unlike the simulation against a canonical specification, the 
abstraction here is a sequential set. The abstraction of
the \Cont operation is therefore given by\vspace{1mm}

{}\hfill
$AbsCont_p  \sdef   {[}\Delta AState, x?:V, r! : \bool |r! = (x? \in S)
]$\hfill {} \vspace{1mm}

\noindent To cope with the non-determinism in the linearization points, yet
allow locality in the proof obligations generated, Derrick et al.
generalise the notion of a status by introducing $INOUT(in, out)$ that
covers a situation in which an operation has {\em potentially}
linearized, where $in$ and $out$ denote the input and output
parameters, respectively. Thus, in this new setting: \vspace{1mm}

\hfill
  $STATUS  \abssynt   IDLE \mid IN\lang V \rang \mid OUT \lang V
  \rang \mid INOUT \lang V \times V \rang $ \hfill {}\vspace{1mm}

  \noindent For example, in the lazy set, process $p$ with status
  $INOUT(3,true)$ denotes a process that is potentially after its
  linearization point, has $3$ as input and will return $true$.

  The proof proceeds by identifying the status for a concrete state
  and a process. For a {\tt contains(x)} operation executed by
  process $p$, given that $cs$ is a concrete state, \vspace{1mm}

\hfill
$\begin{array}[t]{@{}l@{}}
cs.pc(p) = C1 \implies status(cs,p) = IN(x) \\
cs.pc(p) \in \{C2, C3,C4a\} \implies status(cs,p) = INOUT(x, x \in
cs.aux\_S) \\
cs.pc(p) = C4b \implies status(cs,p) = OUT(cs.(val(curr(p))) = x) \\
cs.pc(p) = C5 \implies status(cs,p) = OUT(cs.res(p))
\end{array}$ \hfill {} \vspace{1mm}

\noindent
While executing $C2$, $C3$ or $C4a$, a \Cont operation may now
``change its mind'' about the linearization point and its outcome as
often as necessary.  The proof obligation requires that every change
is justified by the current set representation. In particular, a
process $q$ marking the element that is searched by process $p$ will
change the status of process $p$ executing {\tt contains} to
$false$. This is justified, since the value being searched by $p$ is
also removed from the set representation. A process $q$ adding a cell
with $x$ after $curr(p)$ will change $p$'s status to $true$, which is
justified because $x$ is also added to the abstract set.  

\begin{figure}
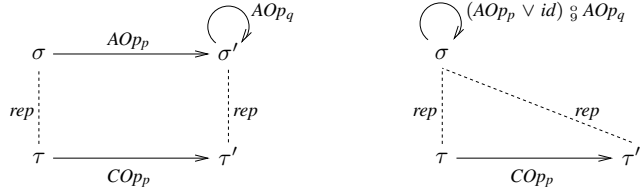

\rule{\textwidth}{1pt}
\begin{center}
    \scalebox{0.9}{\input{simtype3.pspdftex}} \quad \quad\quad \quad\quad \quad
    \scalebox{0.9}{\input{simtype5.pspdftex}}
  \end{center}
  \vspace*{-3.5ex}
  \caption{Additional simulation types}
  \label{fig:simt2}
\rule{\textwidth}{1pt}\end{figure}

To cope with the fact a step in an operation potentially linearizes
those in (several) other operations, two new simulation types are
introduced in addition to those in \reffig{fig:sim} (see \cite{DSW11}
for full details). The left diagram of \reffig{fig:simt2} shows the
case where the execution of operation $COp_p$ definitely sets its own
as well as the linearization point of process $q$ that executes an
operation that does not modify the global state (e.g., a \Cont
operation). The right hand side depicts the case where the abstract
operation of process $p$ is a potential linearization point for $p$
that does not modify the abstract state (e.g,. a \Cont
operation). 

\subsection{Method 3: Augmented states with RGSep}
\label{sec:method-3:-auxiliary}

The method of Vafeiadis also requires substantial changes to cope with
verification of the \Cont operation. In particular, auxiliary
statements that are able to linearize the currently executing \Cont
operations must be introduced to the remove operation. As with the
methods in Sections \ref{sec:method-1:-simulation} and
\ref{sec:method-2:-refinement}, a \Cont operation may linearize
several times before returning, however, the output returned must be
consistent with the state of the queue during the execution of {\tt
  contains}.

The augmented state introduces a further auxiliary variable $OSet$
of type $P \times V \times \bool$, where $(p, v, r) \in OSet$, iff
process $p$ is executing a \Cont operation with input $v$ that has set
its return value to $r$. This requires modification of environment
actions that modify the shared state space. Operations $Lock_p$,
$Unlock_p$, $Add_p$ and $Remove_p$ are as given in
\refsec{sec:auxiliary-variables}. The $Mark_p$ action, which is an
environment action for process $p$ that marks a node must also modify
the abstract set $S$ (as in \refsec{sec:auxiliary-variables}) and the
auxiliary $OSet$. In addition to setting the marked value to $true$
and removing $v$ from the abstract set, the executing process $p$ also
sets the return value of all processes in $C \subseteq OSet$ that
are currently executing a {\tt contains(v)} to $false$, which
linearizes each of the processes in $C$.
\begin{schema}{Mark_p}
  \Delta AugState\\
  n, n1 : Node, 
  B, C : P \times V \times \bool, 
  v : V, r : \bool \ST lvn(n1) = (p, v, n) \land OSet = B \cup C \land
  (\all b : B \st b.2 \neq v) \land (\all c : C \st c.2 = v)
  \\
  lvnm'(n1) = (p, v, n, true) \land 
  S' = S \backslash \{v\} \\ 
  OSet' = B \cup \{(q, v, false) | \exists r \st (q, v, r) \in C\}
\end{schema}
In addition, two environment operations that add and remove triples of
type $P \times V \times \bool$ to/from the auxiliary variable $OSet$
are  introduced. These represent environment processes that start
and complete a \Cont operation.\vspace{1mm}

\hfill
$\begin{array}[t]{rcl}
  AddOut_p & \sdef & [\Delta AugState, 
  v : V, 
  r : \bool | 
  (\all o: OSet \st o.1 \neq p) \land  
  OSet' = OSet \cup \{(p,v,r)\}] \\
  RemOut_p & \sdef & [      \Delta AugState, 
      v : V, 
      r: \bool
      | (p,v,r) \in OSet \land 
      OSet' = OSet \bs\{(p,v,r)\}
]
\end{array}$\hfill{} \vspace{1mm}
\noindent The auxiliary code to the \Add and \Rem operations are as before, but
a {\tt remove(x)} operation must additionally linearize processes in
{\tt OSet} that are executing {\tt contains(x)}. Thus, statement
{\tt R3} is augmented as follows: 

\begin{minipage}{0.95\columnwidth}
    \tt 
    remove(x): 
    
    \ ...

    \ R3: \ \ <n2.mark := true; AbsRes(this) := (x $\in$ S);

    \ \ \ \ \ \ \ \ \textbf{for each} q $\in$ OSet \textbf{do}\ \textbf{if} q.2
    = n2.val \textbf{then} AbsRes(q) := false 
    >  ...
  \end{minipage}

The augmented version of the \Cont operation is given
below\footnote{The presentation in \cite{Vaf07} suffers from a few
  typos, which are confirmed by the proof in \cite{LF13-TR}. In
  particular, the auxiliary code that linearizes itself (in statement
  {\tt C4a}) should only set the abstract result to true if both {\tt
    \textbf{not} res} and {\tt curr.val = e} hold, as opposed to only
  {\tt \textbf{not} res} as indicated in \cite{Vaf07}.}. Like
\cite{Vaf07}, details of the annotation for the proof outline are
elided below, but the interested reader may consult \cite{LF13-TR}.
\begin{center}
  \begin{minipage}{0.99\columnwidth}
    \tt

    contains(x) :

    \ \ \ \ \ <AbsRes(this) := (x $\notin$ S); OSet := OSet $\cup$
    \{this\}>;

    \ C1: curr := Head;

    \ ...
  
  

    C4a: <r1 := curr.marked; 

    \ \ \ \ \ \ AbsRes(this) := (\textbf{not} r1
    \textbf{and} curr.val = x); 

    \ \ \ \ \ \ OutOps := OutOps $\backslash$
    \{this\}> ...
  
  
  \end{minipage}
\end{center}

The augmentation is such that any process $p$ that invokes {\tt
  contains(x)} initially linearizes to $true$ or $false$ depending on
whether or not {\tt x} is in the abstract set, then records itself
in {\tt OSet}. This allows other processes executing {\tt
  remove(x)} to set $p$'s linearization point when {\tt x} is marked
(and logically removed). The linearization point for an execution that
returns $true$ is set at statement {\tt C4a} if $curr(p)$ points to an
unmarked node with value $x$.

\subsection{Discussion}
\label{sec:discussion-2}

The lazy set represents a class of algorithms that can only be
verified by allowing an operation to set the linearization point of
another, and its proof is therefore more involved. The methods we have
considered tackle the problem using seemingly different
techniques. However, translating each proof to a uniform framework, in
this case Z, one can see that the underlying ideas behind the methods
are similar, and experience in verification using one of these methods
can aid in the proof in another. Identifying the linearization points
and understanding the effects of linearization on object at hand
remains the difficult task.  Here, further complications arise because
external operations potentially set the linearization point of the
current operation.

Dongol and Derrick \cite{DD13AVoCS} present a method for verifying
linearizability using an interval-based framework, which aims to
capture the fact that operations like \Cont must only observe the
value being checked as being in the set at some point within its
interval of execution. The logic is able to prove properties of the
form
\begin{eqnarray*}
  beh_p(\Cont(x, true)) \land rely_p & \imp & \Diadot (x \in absSet) \\
  beh_p(\Cont(x, false))  \land rely_p & \imp & \Diadot (x \notin absSet)
\end{eqnarray*}
Here, $beh_p(\Cont(x, true))$ defines an interval-based semantics of
the behaviour of $\Cont(x, true)$ executed by process $p$, $rely_p$ is
an interval predicate that defines the behaviour of the environment of
$p$ and $\Diadot (x \in absSet)$ is an interval predicate that holds
if $x \in absSet$ is true at some point in the given interval. Such
proofs allow one to avoid backward reasoning because the entire
interval of execution is taken into account (\cite{DD13AVoCS}),
however, the interval-based semantics of programs remains complex.

\section{Case study 3: The Herlihy-Wing queue}
\label{sec:case-study-3}
We now discuss the third type of algorithm, where none of the atomic
program statements qualify as linearization points. Instead, execution
of an atomic statement that linearizes an operation depends on future
executions. One such algorithm is the array-based queue by Herlihy and
Wing \cite{Herlihy90}, which we present in \reffig{fig:HWQ}. The
abstract object corresponding to a concrete state cannot be determined
by examining the shared data structure (in this case a shared array)
alone --- one must additionally take into consideration the currently
executing operations and their potential future executions. As these
operations may potentially modify the shared data structure in the
future, each concrete state ends up corresponding to a set of abstract
states.

\begin{figure}
\rule{\textwidth}{1pt}
  \centering 
  \begin{minipage}[t]{0.9\columnwidth}
      \tt \footnotesize enq(lv : V)

      \ E1:\ (k,back) := (back, back+1);  // increment 

      \ E2:\ AR[k]:= lv; // store 

      \ E3:\ \textbf{return}

      \bigskip

      \tt \footnotesize
      
      deq():

      \ D1:\ lback := back; k:=0; lv := null;

      \ D2:\ \textbf{if} k < lback \textbf{goto} D3 \textbf{else goto}
      D1

      \ D3:\ (lv, AR[k]) := (AR[k], lv); // swap

      \ D4:\ \textbf{if} lv != null \textbf{then goto} D6 \textbf{else
        goto} D5

      \ D5:\ k := k + 1; \textbf{goto} D2

      \ D6:\ return(lv)
  \end{minipage}
\caption{The Herlihy-Wing queue}
\label{fig:HWQ}
\rule{\textwidth}{1pt}
\end{figure}

In \reffig{fig:HWQ}, each line corresponds to a single atomic
statement, including for example {\tt D1}, which consists of several
assignments. These operations operate on an infinite array, {\tt AR}
(initially {\tt null} at each index), and use a single shared global
counter, {\tt back} (initially $0$) that points to the end of the
queue.

An enqueue operation ({\tt enq}) atomically increments {\tt back}
(line {\tt E1}) and stores the old value of {\tt back} locally in a
variable {\tt k}. Thus executing {\tt E1} allows the executing process
to reserve the index of {\tt back} before the increment as the
location at which the enqueue will be performed. The enqueued value is
stored at {\tt E2}. A dequeue operation ({\tt deq}) stores {\tt back}
locally in {\tt lback}, then traverses {\tt AR} from the front (i.e.,
from index $0$) using {\tt k}. As it traverses {\tt AR}, it swaps the
value of {\tt AR} at {\tt k} with {\tt null} ({\tt D3}). If a non-null
element is encountered ({\tt D4}), then this value is returned as the
head of the queue. If the traversal reaches {\tt lback} (i.e., the
local copy of {\tt back} read at line {\tt D1}) and a non-null element
has not been found, then the operation restarts. Note that {\tt deq}
is \emph{partial} \cite{Herlihy90} in that it does not terminate if
{\tt AR} is {\tt null} at every index. In particular, a dequeue only
terminates if it returns a value from the queue.

To see why verifying linearizability of the algorithm is difficult, we
show that neither {\tt E1} nor {\tt E2} qualify as a valid
linearization points for {\tt enq}. It is straightforward to derive a
similar counter example for {\tt E3}. Suppose {\tt E1} is picked as
the linearization point and consider the following complete execution,
where $p,q,r \in P$. Assume $p$ and $q$ enqueue $v_1$ and $v_2$,
respectively.
\begin{eqnarray}
  \label{eq:6}
  \langle {\tt E1}_p,  {\tt E1}_q, {\tt E2}_q, {\tt D1}_r, {\tt D2}_r,
  {\tt D3}_r, {\tt D4}_r, {\tt D5}_r, 
  {\tt D2}_r, {\tt D3}_r, {\tt D4}_r, {\tt D6}_r, {\tt E2}_p, {\tt E3}_q, {\tt
    E3}_p \rangle 
\end{eqnarray}
Although ${\tt E1}_p$ is executed before ${\tt E1}_q$, the dequeue
operation returns $v_2$ before $v_1$, contradicting FIFO ordering, and
hence, {\tt E1} cannot be a linearization point. Now suppose {\tt E2}
is picked as the linearization point and consider the following
complete execution:
\begin{eqnarray}
  \label{eq:7}
  \langle {\tt E1}_p,  {\tt E1}_q, {\tt D1}_r, {\tt D2}_r,
  {\tt D3}_r, {\tt D4}_r, {\tt D5}_r, {\tt E2}_p, {\tt E2}_q, {\tt D2}_r,
  {\tt D3}_r, {\tt D4}_r, {\tt D6}_r, {\tt E3}_q, {\tt
    E3}_p \rangle 
\end{eqnarray}
Now, ${\tt E2}_p$ is executed before ${\tt E2}_q$, but {\tt deq}
returns $v_2$ before $v_1$ has been dequeued. 

The histories corresponding to both executions are however,
linearizable because the operation calls ${\tt enq}_p$, ${\tt enq}_q$
and ${\tt deq}_r$ overlap, allowing their effects to occur in any
order. In particular, both \refeqn{eq:6} and \refeqn{eq:7} correspond
to history
\begin{eqnarray*}
  \langle enq_p^I(v_1),  enq_q^I(v_2), deq_r^I, deq_r^R(v_2), enq_q^R,
  enq_p^R\rangle 
\end{eqnarray*}
which is linearizable.

Aside from the proof sketch in Herlihy/Wing's original paper
\cite{Herlihy90}, there are two known formal proofs of
linearizability: Schellhorn \etal \cite{SWD12,SDW14} (which uses
backwards simulation) and \cite{HSV13} (which decomposes the problem
into several \emph{aspects}). Henzinger et al's main ordering property
requires use of prophecy variables, and hence must perform reasoning
similar to backward simulation.

Backward simulation and prophecy variables are known to be equivalent
formulations, that allow the future non-determinism to be taken into
account \cite{deRoever98}. Both allow one to capture the fact that in
order to decide whether the enqueue operation has taken effect, one
must consider the state of all currently executing operations

\subsection{Method 1: Backward simulation proofs}
\label{sec:method-1:-backward}
Schellhorn \etal \cite{SDW14} have shown that backward simulation is
sufficient for proving linearizability, i.e., backward simulation is
complete. These methods however, do not show \emph{how} such a
simulation relation may be constructed, and hence, creativity is
required on the part of the verifier to develop the correct simulation
relation. As already discussed, each concrete state corresponds to
multiple abstract queues depending on the states of the executing
operations. Schellhorn et al's approach is to encode, within the
simulation relation, all possible ways in which the currently
executing {\tt enq} operations can complete, as well as all possible
ways in which these could be dequeued by. To this end, they construct
a so-called \emph{observation tree}. In effect, this constructs the
\emph{set} of all possible queues that could relate to the current
concrete queue based on the state of {\tt AR} and the pending
concurrent operations. The proof methods build on previous work on
potential linearization points (\refsec{sec:method-2:-refinement}),
the difference here is that linearizing external operations modifies
the data structure in question.

For example, statement {\tt E1} of {\tt enq} is a potential
linearization point, and hence, one must perform case analysis to
check whether or not its execution linearizes the currently executing
enqueue operation. The non-linearizing cases are straightforward as
one must only check that the set of queues from the post-state are the
same as those in the pre-state. For the linearizing case, there must
be some abstract queue related to the concrete post-state for which
the element being enqueued is at the tail of some abstract queue
related to the pre-state. Proving this is further complicated by the
fact that an {\tt enq} operation call executing {\tt E1} may
`overtake' other {\tt enq} operation calls that executed {\tt E1}
earlier (and hence have a lower local value of {\tt k}), causing the
effect of a latter execution of {\tt E1} to occur first. In fact,
depending on the configuration of operation calls in the concrete
state, executing {\tt E1} may even overtake other {\tt enq} operation
calls that have executed {\tt E2}.


The full argument is rather complex, and hence, we do not present
further details of this verification here. Instead we ask the
interested reader to consult \cite{SDW14}. We note however, that their
proofs are fully mechanised using the KIV theorem prover.

\subsection{Method 2: Aspect-oriented proofs}
\label{sec:method-2:-aspect}
A second proof of the Herlihy/Wing queue is given by Henzinger \etal
\cite{HSV13}, who define a set of \emph{aspects} that characterise the
behaviour of a queue and show that Herlihy/Wing's queue satisfies
these aspects. In particular, the following aspects are required of a
FIFO queue:
\begin{description}
\item [\sf VFresh] A dequeue event returning a value not inserted by any
  enqueue event.
\item [\sf VRepet] Two dequeue events returning the value
  inserted by the same enqueue event.
\item [\sf VOrd] Two ordered dequeue events returning values inserted
  by enqueue events in the inverse order.
\item [\sf VWit] A dequeue event returning empty even though the queue
  is never logically empty during the execution of the dequeue event.
\end{description}
These aspects are only shown to be necessary and sufficient for proving
linearizability if the implementation is \emph{purely-blocking},
meaning that from any reachable state, any pending operation, if run
in isolation will either terminate or its entire execution does not
modify the global state.

For the Herlihy-Wing queue, {\sf VWit} is irrelevant as the dequeue
loop only terminates if a non-null element is read, i.e., it never
returns empty. Both aspects {\sf VFresh} and {\sf VRepet} are
straightforward to check. {\sf VOrd}, however is more involved as it
must reason about potential reordering of {\tt enq} operation
encountered by \cite{SDW14}. Aspect {\sf VOrd} is reformulated as {\sf
  POrd}, which states the following.
\begin{quote}
  Fix a value $v_2$ and consider a history $c$ where every method call
  enqueuing $v_2$ is preceded by some method call enqueuing some
  different value $v_1$ and there are no {\tt deq()} calls returning
  $v_1$ (there may be arbitrarily many concurrent {\tt enq()} and {\tt
    deq()} calls enqueuing or dequeuing other values). The goal is to
  show that in this history, no {\tt deq()} return $v_2$. \cite{HSV13}
\end{quote}
In other words, if an ordering of values $v_1$ and $v_2$ in a history
$c$ has been decided so that the enqueue of $v_1$ precedes the enqueue
of $v_2$, and no dequeue operation calls return the first value $v_1$,
then there are no dequeue operations that dequeue the second. 

The proof {\sf POrd} for the Herlihy/Wing queue requires the use of
\emph{prophecy variables} that allow dequeue operations to `guess' the
value that they will dequeue. Assertions on prophecy variables are
encoded as assertions within the program code, then verification
proceeds by showing that these guesses are correct. Again, we leave
out the full details of the proof method, and ask the interested
reader to consult \cite{HSV13}.

\subsection{Discussion}

The Herlihy-Wing queue represents a class of algorithms that can only
be proved linearizable by considering the future behaviours of the
currently executing operation calls, further complicated by the
potential for these current operations to modify the data structure at
hand. Reasoning must therefore appeal to backward simulation or by
prophecy variables.

Schellhorn \cite{SDW14} use a backward simulation consisting of a
monolithic simulation relation that captures all possible future
behaviours at the abstract level. The method has been show to be
complete for verifying linearizability, however, developing and
verifying such a simulation relation is a complex task. The
aspect-oriented proof method has decomposition of a linearizability
proof for purely blocking algorithms into simpler aspects that are (in
theory) easier to verify \cite{HSV13}. However, it is currently not
clear whether every data structure can be decomposed into aspects, and
whether the method does truly simplify proof of the most difficult
portions.

These are not the only method capable of handling future linearization
points --- two other methods, both based on backward simulation, could
be applied to verify the Herlihy-Wing queue. We have not presented a
detailed comparison here as they have not verified the Herlihy-Wing
queue (i.e., we do not attempt a proof using their methods
ourselves). Groves et al's backward simulations against canonical
automata can cope with future linearization points
\cite{Doherty04DCAS}. Tofan \etal \cite{TSR14} have continued to
improve the simulation-based methods
(\refsec{sec:method-1:-backward}), and incorporated the core theory
into a interval-based rely/guarantee framework. Here, linearizability
is re-encoded using \emph{possibilities}, which describe the orders of
completions of pending operation calls. Their methods have been
applied to verify correctness of an array-based multiset with insert,
delete, and lookup operations. An interesting aspect of this algorithm
is that it is possible for a lookup of an element $x$ to return
$false$ even if the element $x$ is in the array in all concrete states
throughout the execution of the lookup operation. Their methods have
been linked to the completeness results of Schellhorn \etal
\cite{SDW14}.

\section{Conclusions}
\label{sec:conclusions}

There has been remarkable progress since Herlihy and Wing's original
paper on linearizability \cite{Herlihy90}, and with the increasing
necessity for concurrency, this trend is set to continue. The basic
idea behind linearizability is simple, yet it provides a robust
consistency condition applicable to a large number of algorithms, and
in some cases precisely captures the meaning of atomicity
\cite{Ray13}. Linearizability is \emph{compositional} in the sense
that a set of objects is linearizable if each object in the set is
linearizable \cite{HS08,Herlihy90}, making it an appealing
property. Besides shared variable concurrent objects, linearizability
has also been applied to distributed systems \cite{Bir92}, databases
\cite{RC92} and fault-tolerant systems \cite{GS96}.

This paper considered verification of linearizability, and the
associated proof methods that have been developed for it in the
context of concurrent objects. Necessity of such proofs is alluded to
by the subtleties in the behaviours of the algorithms that implement
concurrent objects, and by the fact that it's errors have been found
in algorithms that were previously believed to be correct
\cite{Doherty04DCAS,CG05}. Current proof techniques continue to
struggle with the scalability and as a result, only a handful of
fine-grained algorithms have been formally verified to be
linearizable. The longest fully verified algorithm (in terms of lines
of code) is the Snark algorithm \cite{Doherty04DCAS}. However, number
of lines of code is not an indicator of complexity, with even simple
algorithms like Herlihy and Wing's queue \cite{Herlihy90} posing
immense challenges \cite{SWD12,SDW14,HSV13} due to the fact that
future behaviour must be considered.

Our survey has aimed to answer the questions that were posed in
\refsec{sec:introduction}. We now return to these to discuss
concluding remarks. \vspace{1mm}

\noindent 
{\it Locality of the proof method.} Each of the methods we've
considered enable localised reasoning, only requiring the behaviour of
a single process to be considered. However, interference must be
accounted for in the invariants and refinement relations generated,
complicating each verification step. Namely, one must show that an
invariant holds locally and is preserved by the each step of an
arbitrarily chosen process, and that it holds in the presence of
interference from other processes.
\vspace{1mm}

\noindent {\it Compositionality of the proof method.} Some methods
have incorporated Jones-style rely/guarantee reasoning into their
respective frameworks (e.g., RGSep and RGITL), allowing potential
interference from the environment to be captured abstractly by a rely
condition. An additional step of reasoning is required to show that
the rely condition is indeed an abstraction of the potential
interference, but once this is done, a reduction in the proof load is
achieved via a reduction in the number of cases that must be
considered.  \vspace{1mm}

\noindent {\it Contribution of the underlying framework.} None of the
existing frameworks thus far provide a silver bullet for
linearizability verification. Identification of the linearization
points and appropriate representation relations remain the difficult
aspects of a proof. If the verifier believes an algorithm to have
fixed linearization points, then it would be fruitful to attempt an
initial verification using a tool such as the one provided by
Vafeiadis \cite{Vaf10}. For more complex algorithms, using a setup
such as the one provided by Colvin \etal \cite{CGLM06} would allow
invariants to be model checked prior to verification. On the other
hand, Derrick \etal \cite{DSW11TOPLAS} have developed a systematic
method for constructing representation relations, invariants and
interference freedom conditions as well as proof obligations that
enables process-local verification. Techniques specific to certain
implementations (e.g., the Hindsight Lemma, aspect-oriented
verification) enable some decomposition possibilities, but have not
been generalised to cope with arbitrary implementations.  \vspace{1mm}

\noindent
{\it Algorithms verified.} A survey of these has been given in
\refsec{sec:degrees-difficulty}. There exist several other algorithms
in the literature whose linearizability has been conjectured, but not
yet formally verified has not yet been performed. For the frameworks
we've studied, the number of algorithms verified is however not a
measure of it capabilities; rather it is whether the framework can
handle complex algorithms with future linearization points such as the
Herlihy/Wing queue. 

The verifications thus far, have only considered linear (flat) data
structures. Recently, more challenging structures such as SkipTries
\cite{OS13} and binary search trees \cite{CDT14} have been
developed. Their linearizability has been informally argued, but not
mechanically verified. It is not easy to know exactly how the proof
complexity increases for such data structures, however, the complex
nature of the underlying algorithm and the abstract representations
suggest that the proofs will also be more complex.  \vspace{1mm}

\noindent
{\it Mechanisation.} Many of the methods described in this paper
have additional tool support that support mechanical validation of the
proof obligations, reducing the potential for human error. In some
cases, automation has been achieved, reducing human effort, but these
are currently only successful for algorithms with fixed linearization
points and a limited number of algorithms with external linearization
points. \vspace{1mm}

\noindent {\it Completeness.} Completeness of a proof method is
clearly a desirable quality --- especially for proofs of
linearizability, which require considerable effort. Backwards
simulation alone is known to be complete for verifying linearizability
against an abstract sequential specification
\cite{SWD12,SDW14}. Furthermore, a combination of forwards and
backwards simulations is known to be complete for data refinement
\cite{Lyn96-DA,deRoever98}, and combining auxiliary and prophecy
variables is known to be complete for reasoning about past and future
behaviour \cite{AL91-existence}. Completeness of a method does not
guarantee simpler proofs, as evidenced by the maximal backwards
simulation constructed by Schellhorn et al.~\cite{SWD12,SDW14} to
prove linearizability.  The completeness results Schellhorn \etal
\cite{SWD12,SDW14} show that by using the global theory any
linearizable algorithm can be proved correct. This shows that for
every linearizable object, a backward simulation in between abstract
and concrete specification can be found. This result does, however,
not directly give one a way of constructing this backward
simulation. This is common to all completeness results: they state the
existence of a proof within a particular framework, but not the way of
finding this proof.  That such proofs can for individual instances
indeed be found, is exemplified by the highly non-trivial case study
of the paper.

\paragraph{Model checking}
An important strand of research is model checking, which does not
always prove linearizability, but can be used to check invariants
needed to verify linearizability. This paper has focused on
verification methods, and hence, a detailed comparison of model
checking methods have been elided. However, like Colvin \etal
\cite{CGLM06}, we believe model checking can play a complementary role
in verification, allowing invariants to be model checked prior to
verification to provide assurances that they can be proved
correct. Methods for model checking linearizability may be found in
\cite{VYY09,Fri13,LCLS09,L0L0ZD13}; a comparison of these techniques
is beyond the scope of this survey.

\paragraph{Progress properties}
In many applications, one must often consider the progress properties
that an algorithm guarantees. Here, like safety, several different
types of progress conditions have been identified such as
\emph{starvation freedom}, \emph{wait freedom}, \emph{lock freedom}
and \emph{obstruction freedom} (see
\cite{HS08,HS11,Don09,Don06ICFEM,TBSR10,LiangHFS13,GotsmanCPV09}). Progress
properties are not the main focus of this paper, and hence, discussion
of methods for verifying them have been elided. Nevertheless, they
remain an important property to consider when developing algorithms.

\paragraph{Relaxing linearizability} 
The increasing popularity of multicore/multiprocess architectures, has
led to an increasing necessity for highly optimised algorithms. Here,
researchers are questioning whether linearizability is itself causing
sequential bottlenecks, which in turn affects performance. Due to
Amdahl's Law, it is known that if only 10\% of a program's code
remains sequential, then one can achieve at best a five-fold speedup
on a 10-core machine, meaning at least half of the machine's
capability is wasted \cite{Sha11,MS04}. As a result, Shavit
\cite{Sha11} predicts future systems will trend towards more relaxed
notions of correctness.

Several conditions weaker than linearizability have been defined to
allow greater flexibility in an implementation, e.g.,
\emph{quasi-linearizability} \cite{AfekKY10},
\emph{$k$-linearizability} \cite{HKPSS13}, \emph{eventual consistency}
\cite{SK09}.  Part of the problem is that linearizability insists on
\emph{sequential consistency} \cite{Lam97,HS08}, i.e., that the order
of events within a process is maintained. However, modern processors
use local caches for efficiency, and hence, are not sequentially
consistent. Instead, they only implement \emph{weak memory models}
that allow memory instructions to be reordered in a restricted manner
\cite{AG96}. Shavit \cite{Sha11} purports \emph{quiescent consistency}
as the correctness criteria for objects in the multicore age, which
only requires the real-time order of operation calls to be maintained
when the calls are separated by a period of quiescence (which is a
period without any pending operation invocations). Unlike
linearizability, quiescent consistency does not imply sequential
consistency, and hence, is applicable to weak memory models
\cite{SDD14}. As quiescent consistency is weak condition, more recent
work has consider quantitative relaxations to bridge the gap between
linearizability and quiescent consistency \cite{JR14}.

Weakening correctness conditions however, does not mean that the
algorithms become easier to verify and furthermore methods for
verifying linearizability can be ported to weaker conditions (e.g.,
see \cite{DerrickDSTTW14}). Therefore, techniques for simplifying
linearizability proofs will not be in vain if in the future weaker
conditions become the accepted standard.

\paragraph{Future directions}
Despite the numerous advances in verification methodologies, formal
correctness proofs of concurrent algorithms in a scalable manner
remains an open problem. This in turn affects verification of specific
properties such as linearizability. The rate at which new algorithms
are developed far outpace the rate at which these algorithms are
formally verified.  However, as concurrent implementations become
increasingly prevalent within programming libraries (e.g.,
\texttt{java.util.concurrent}) the need for formal verification
remains important.

So what will future algorithms look like? To reduce sequential
bottlenecks, there is no doubt that concurrent objects of the future
will continue to become more sophisticated with more subtle
(architecture-specific) optimisations becoming prevalent. Proving
linearizability of such algorithms will almost certainly require
consideration of some aspect of future behaviour. It is therefore
imperative that verification techniques that are able to handle this
complex class of algorithms continue to be improved. The frameworks
themselves must continue to integrate the various methods for proof
decomposition (e.g., \refsec{sec:meth-proof-decomp}). For example,
Tofan \etal \cite{TSR14} have developed a framework that combines
interval temporal logic, rely/guarantee and simulation proofs. Further
simplifications could be achieved by extending the framework with
aspects of separation logic. In some cases, decomposition of a proof
into stages, e.g., using reduction, or interval-based abstraction has
been useful, where the decomposition not only reduces the number of
statements that must be considered, but also transfers the algorithm
from a proof that requires consideration of external linearization
points to a proof with fixed linearization points. Until a scalable
generic solution is found, it is worthwhile pursuing problem-specific
approaches (e.g., \cite{HSV13,DGH13}).

Another avenue of work is proof modularisation. To explain this,
consider the elimination queue \cite{MoirNSS05}, which embeds an
elimination mechanism (implemented as an array) on top of the queue by
Michael and Scott \cite{Michael96} (with some modifications). Although
linearizability of Michael and Scott's queue is well studied, current
techniques require the entire elimination queue data structure to be
verified from scratch. Development of modular proof techniques would
enable linearizability proofs to be lifted from low-level data
structures to more complex (optimised) versions. New results such as
parameterised linearizablity \cite{CGY14} suggest that modular
concurrent objects and associated proof techniques will continue to
evolve.

\bibliographystyle{plain}
\bibliography{ls}

\end{document}